\documentclass[floatfix,aps,prd,eqsecnum,showpacs,amsmath,nofootinbib,preprintnumbers]{revtex4}

\usepackage{graphicx}

\usepackage{hyperref}
\hypersetup{colorlinks=true,linkcolor=blue,pdfstartview=XYZ}

\def\laq{~\raise 0.4ex\hbox{$<$}\kern -0.8em\lower 0.62ex\hbox{$\sim$}~}
\def\gaq{~\raise 0.4ex\hbox{$>$}\kern -0.7em\lower 0.62ex\hbox{$\sim$}~}

\def\beq{\begin{equation}}
\def\eeq{\end{equation}}
\def\bea{\begin{eqnarray}}
\def\eea{\end{eqnarray}}
\def\bean{\begin{eqnarray*}}
\def\eean{\end{eqnarray*}}

\def\re {\rangle}

\def \pa {\partial}
\def \ra {\rightarrow}
\def \ti {\widetilde}
\def \la {\lambda}

\def \La {\Lambda}
\def \Da {\Delta}
\def \da {\delta}

\def \ga {\gamma}
\def \sg {\sigma}
\def \Sg {\Sigma}
\def \da {\delta}

\def \Om {\Omega}

    \def\be{\begin{equation}}
    \def\ee{\end{equation}}
    \def\ba{\begin{eqnarray}}
    \def\ea{\end{eqnarray}}

    \def\e{\mbox{e}}

\newcommand{\eq}{\begin{equation}}
\newcommand{\eqx}{\end{equation}}
\newcommand{\eqn}{\begin{eqnarray}}
\newcommand{\eqnx}{\end{eqnarray}}

\newcommand{\Ups}{\Upsilon}

\newcommand{\rref}[1]{(\ref{#1})}

\newcommand{\lla}{\left\langle}
\newcommand{\rra}{\right\rangle}

\begin{document}

\preprint{BA-TH/650-12}
\preprint{CERN-PH-TH/2012-018}
\preprint{LPTENS-11/46}

\title{Backreaction on the luminosity-redshift  relation \\ from gauge invariant light-cone averaging}

\author{I. Ben-Dayan$^{1,2}$, M. Gasperini$^{3,4}$, G. Marozzi$^{5}$, F. Nugier$^{6}$ and G. Veneziano$^{5,7}$}

\affiliation{$^1$Canadian Institute for Theoretical Astrophysics, 60 St
George, Toronto ON, M5S 3H8\\
$^{2}$ Perimeter Institute for Theoretical Physics,
Waterloo, Ontario N2L 2Y5, Canada\\
$^{3}$Dipartimento di Fisica, Universit\`{a} di Bari, Via G. Amendola
173, 70126 Bari, Italy\\
$^{4}$Istituto Nazionale di Fisica Nucleare, Sezione di Bari, Bari, Italy\\
 $^{5}$ Coll\`ege de France, 11 Place M. Berthelot, 75005 Paris, 
 France\\
$^{6}$ Laboratoire de Physique Th\'eorique de l'\'Ecole Normale Sup\'erieure, CNRS UMR 8549, 24 Rue Lhomond, 75005 Paris, France\\
$^{7}$CERN, Theory Unit, Physics Department, \\ CH-1211 Geneva 23, Switzerland}


\begin{abstract}

Using a recently proposed gauge invariant formulation of light-cone averaging, together with adapted ``geodesic light-cone" coordinates, we show how an ``induced backreaction" effect  emerges, in general, from correlated fluctuations in the luminosity distance and  covariant integration measure. Considering  a realistic stochastic spectrum of inhomogeneities of primordial (inflationary) origin we find that both the induced backreaction on the luminosity-redshift relation and the dispersion are larger than na\"ively expected. On the other hand the former, at least to leading order and in the linear perturbative regime,  cannot account by itself for the observed effects of dark energy at large-redshifts. A full second-order calculation, or even better a reliable estimate of contributions from the non-linear regime, appears to be necessary before firm conclusions on the correct interpretation of the data can be drawn.

\end{abstract}

\vspace {1cm}~


\maketitle

\section {Introduction}
\label{Sec1}
\setcounter{equation}{0}

The so-called concordance (or $\Lambda$CDM) model, based on a suitable combination of dark matter, dark energy and baryons for an overall critical density, has become the reference paradigm for the late  -- i.e. post-equality epoch -- evolution of our Universe (see e.g. \cite{pdg}). It accounts equally well for the CMB data, the Large Scale Structure and, even more significantly, for the 
supernovae data in terms of a cosmic acceleration \cite{nobel}.

Strictly speaking these three tests of the concordance model are not at the same level of theoretical rigor. While the first two have to do, by definition, with the inhomogeneities present in our Universe, the third is based on an ideal homogeneous and isotropic Friedmann-Lema\^itre-Robertson-Walker (FLRW) geometry. It is clear that a better treatment of cosmic acceleration should take inhomogeneities into account, at least in an average statistical sense. Only when this is done we can establish in a convincing way whether $\Lambda$CDM gives a simultaneous consistent description of the above-mentioned  body of cosmological data.

This realization has led to a vast literature about averaging cosmological observables in realistic inhomogeneous cosmologies (see e.g. \cite{review} 
for recent reviews). The conclusions, however, are still rather controversial: according to some authors \cite{3ref} present inhomogeneities might explain, by themselves, cosmic acceleration without any need for dark-energy contributions; according to others \cite{negligible} the effect of inhomogeneities is, instead, completely negligible. The truth may lie somewhere in between, in the sense that a quantitative understanding of inhomogeneities effects could be important in order to put precise constraints on dark-energy parameters, such as the critical fraction of dark-energy density, $\Omega_{\La}$, and the time evolution of its effective equation of state, $w_{\La}(z)$. 

In the first papers studying the dynamical  effects of averaging,  the problem was approached mainly following Buchert's prescriptions \cite{1}, namely averaging inhomogeneities over spacelike hypersurfaces and computing the ensuing ``backreaction" on the averaged geometry. Nonetheless it is clear that a proper treatment of cosmic acceleration -- which is indeed revealed through the experimental study of the luminosity-distance to redshift relation \cite{nobel} -- needs to consider the backreaction of averaged inhomogeneities along our past light cone, which is a null three-dimensional hypersurface.

While a number of papers have emphasized the importance of light-cone averaging (see e.g. \cite{Maartens1,Maartens2,3}), a fully covariant and gauge invariant formulation of such a procedure was only given recently \cite{GMNV}, by generalizing to null hypersurfaces  an analogous prescription previously derived for spacelike hypersurfaces \cite{GMV1, GMV2}. Actually, it turns out that the physically meaningful (covariant and gauge invariant) average of a scalar over a null hypersurface reduces to averaging over an appropriate two-dimensional surface a scalar object which is non local, as the integrand itself contains integrals along lightlike geodesic curves lying on the given null hypersurface. Nonetheless, we shall simply refer to this procedure as ``light-cone averaging''  \cite{GMNV}.

The aim of this paper is to apply such an averaging procedure to the luminosity distance of a light source lying on our past light-cone, in order to compute the possible backreaction of inhomogeneities on the luminosity-redshift relation. We will show the emergence of an effect  -- called hereafter  {\em ``induced backreaction"} -- which arises from a generic correlation between the inhomogeneities present in the  variable we want to average (e.g. the luminosity distance) and those appearing  in the  covariant integration measure\footnote{As discussed in \cite{GMNV}, this integration measure will also induce ``backreaction" terms of the type usually discussed in the literature \cite{1}, namely terms that arise  from (generalized) commutation rules between differential operators and averaging integrals.}. We stress immediately that our  induced backreaction accounts only for a part of the total effect of inhomogeneities. As we shall discuss in detail below, the leading-order
 induced backreaction can be computed in terms of linear perturbation theory, while a complete leading-order calculation (left to future work) would require perturbation theory up to second order.

We will use a simple phenomenological model of inhomogeneous geometry based on a spatially flat FLRW metric which includes, to first order, scalar perturbations of primordial (inflationary) origin. Such perturbations are often  conveniently  parametrized in the longitudinal (or Newtonian) gauge \cite{Muk}. However, we will take  advantage of the gauge invariance of our formalism to use an adapted system of coordinates defining the so-called ``geodesic light-cone'' (GLC) gauge introduced in \cite{GMNV}. In that gauge the light-cone averages of luminosity distance and redshift take very simple, exact expressions that keep all the required degrees of freedom for being applicable to general geometries.   

By further taking the {\em ensemble} average of the inhomogeneous terms, and using the stochastic properties of our model of perturbations,  we will compute
to leading order the induced backreaction on the luminosity distance, and the corresponding variance.
 If we limit ourselves to first order computations in the range of scales where  linear perturbation theory is reliable,  we find that the induced (second-order) corrections to the luminosity-redshift relation of the homogeneous CDM model, although much bigger than one could have na\"ively expected, are not large enough to mimic a sizable fraction of $\Omega_{\Lambda}$.  However, our formalism suggests that other second-order corrections could be even larger, thus confirming the importance of  performing a full  second-order calculation or, even better, a reliable calculation in the non-linear (short scale) regime.

The paper is organized as follows. 
In Sect. \ref{Sec2} we recall definition and basic properties of GLC coordinates  and give exact, non-perturbative expressions for the redshift, the luminosity distance and the light-cone average of the latter at constant redshift. In Sect. \ref{Sec3} we give the linear order transformations expressing  the metric of a perturbed FLRW geometry in GLC form, starting from the more commonly used Newtonian gauge. We then express the luminosity distance as a function of the redshift and of the angular position of the source, to first order in the given perturbed geometry. In Sect. \ref{Sec4} we take the light-cone average of the computed luminosity distance and we show how backreaction effects automatically emerge, in general,  from correlations between the  averaged variable and the covariant integration measure. We explicitly define such an  induced backreaction
and the cosmic variance, to leading order, taking into account the stochastic properties of our model of perturbations and thus including the appropriate {\em ensemble} average of the inhomogeneous corrections. In Sect. \ref{Sec5} we implement analytical calculations of the induced backreaction terms and of the dispersion, and in Sect. \ref{Sec6} we present the corresponding numerical results for a specific spectrum of primordial scalar perturbations and for a given model of transfer function.  Our main conclusions are finally summarized in Sect. \ref{Sec7}. We also present, in Appendix A, a sample of the analytic manipulations needed for reducing our (light-cone plus stochastic) averages to the form of a one-dimensional integral over the wavenumber $k$ labelling the scalar perturbation spectrum. 


\section{The GLC gauge and  the averaged luminosity-redshift relation}
\label{Sec2}
\setcounter{equation}{0}

\subsection{Reminder of the GLC gauge and of its main properties}
\label{Sec21}

Our purpose, in this paper, is to compute the light-cone average of the luminosity distance as a function of the redshift by applying the gauge invariant procedure introduced in \cite{GMNV}. One of the main virtues of using a gauge invariant formalism is the freedom of choosing a gauge particularly adapted to the problem at hand. In the case of spacelike averaging, for instance,  a convenient coordinate system corresponds to a gauge where the averaging hypersurfaces are identified with constant-time hypersurfaces. 
In many applications the averaging hypersurfaces are chosen indeed as the ones associated to a class of geodesic observers  corresponding to constant values of the time parameter $t$ of the synchronous gauge (see e.g.  \cite{Marozzi}).

Similarly, for light-cone averages, it is convenient to identify the null hypersurfaces with those on which a null coordinate takes constant values. For this reason  we have  introduced in \cite{GMNV}  an adapted system of coordinates -- defining what  we have called a  ``geodesic light-cone'' (GLC) gauge -- where the averaging prescription greatly simplifies, while keeping all the required degrees of freedom for applications to general geometries. 
Furthermore we are also able to identify the timelike coordinate of the GLC gauge with the cosmic  time $t$ of the synchronous gauge. As a consequence we can easily introduce a family of geodesic reference observers which exactly coincide with the static ones of the synchronous gauge.

Let us first recall, as discussed in \cite{GMNV}, that the coordinates $x^\mu= (\tau, w, \ti{\theta}^a)$, $a=1,2$, specifying the metric in the GLC gauge correspond to a complete gauge fixing of the so-called observational coordinates,  defined e.g in  \cite{Maartens1,Maartens3,Clarkson}. The GLC metric depends on six arbitrary functions 
($\Ups$, a two-dimensional ``vector" $U^a$ and a symmetric matrix $\gamma_{ab}$),
and its line element takes the form 
\bea
\label{LCmetric}
ds_{GLC}^2 = \Ups^2 dw^2-2\Ups dw d\tau+\gamma_{ab}(d\ti{\theta}^a-U^a dw)(d\ti\theta^b-U^b dw).
 \eea
In matrix form,  the metric and its inverse are then given by:
\eq
\label{GLCmetric}
g^{GLC}_{\mu\nu} =
\left(
\begin{array}{ccc}
0 & - \Ups &  \vec{0} \\
-\Ups & \Ups^2 + U^2 & -U_b \\
\vec0^{\,T}  &-U_a^T  & \gamma_{ab} \\
\end{array}
\right),~~ ~~~~~
g_{GLC}^{\mu\nu} =
\left(
\begin{array}{ccc}
-1 & -\Ups^{-1} & -U^b/\Ups \\
-\Ups^{-1} & 0 & \vec{0} \\
-(U^a)^T/\Ups & \vec{0}^{\, T} & \gamma^{ab}
\end{array}
\right) ~,
\eqx
where $\vec 0=(0,0)$, $U_b= (U_1, U_2)$, while the $2 \times 2$ matrices 
$\gamma_{ab}$ and  $\gamma^{ab}$ lower and  raise  the two-dimensional indices. Clearly $w$ is a null coordinate (i.e.  $\pa_\mu w \pa^\mu w=0$), and a past light-cone hypersurface is specified by the condition $w=$ const. We can also easily check that $\partial_{\mu} \tau$ defines a geodesic flow, i.e. that
$\left( \pa^\nu \tau\right) \nabla_\nu \left( \pa_\mu \tau\right) = 0$
(as a consequence of the relation $g^{\tau \tau} = -1$). 

In the limiting case  of  a spatially flat FLRW geometry, with scale factor $a$, cosmic time $t$, and conformal time parameter $\eta$ such that $d\eta= dt/a$, the transformations to the GLC coordinates and the  meaning of the new metric components are easily found as follows \cite{GMNV}:
\bea
&&
\tau=t,  ~~~~~~~~~~~~~~~ w= r+\eta,~~~~~~~~~~~
\Ups = a(t),
\nonumber \\ &&
U^a=0, ~~~~~~  \gamma_{ab} d \theta^a d\theta^b = a^2(t) r^2 (d \theta^2 +\sin^2 \theta d\phi^2).
\label{FR}
\eea
Even though we will be mainly using the GLC gauge for a perturbed FLRW metric in the Newtonian gauge,  it is important to stress that the equality between the coordinate $\tau$ and the proper time $t$ of the synchronous gauge holds at the exact, non perturbative level: it is always possible, in fact, to choose the GLC coordinates in such a way that $\tau$ and $t$ are identified like in the above  FLRW limit.

In order to illustrate this point let us consider an arbitrary space-time metric written in the synchronous gauge (with coordinates $X^{\mu} = (t, X^i)$, $i,j= 1,2,3$), where the line element takes the form:
\beq
ds_{SG}^2=-dt^2+h_{ij}dX^i dX^j.
 \eeq
Let us impose the condition  $t=\tau$, and check whether we run into any contradiction with the exact metric transformation
\be
g_{SG}^{\rho\sigma}(X)=\frac{\partial X^\rho}{\partial x^\mu} \frac{\partial X^\sigma}{\partial x^\nu} g_{GLC}^{\mu\nu}(x) .\label{EqBetweenGauges2}
 \ee
Using Eq. (\ref{GLCmetric}) for the GLC metric, and considering the transformation for the $g_{SG}^{t \mu}$ components, we then obtain the conditions
 \be
 g_{SG}^{t\mu}=\{-1,\vec{0}\}=-\left[\partial_{\tau}+\Ups^{-1}(\partial_w+U^a\partial_a)\right]X^{\mu}= u^{\nu}\partial_{\nu}X^{\mu}= \frac{dX^{\mu}}{d\lambda},
 \ee
 where  $u_{\mu} = - \partial_{\mu}\tau$ is the four-velocity of
the geodesic GLC observer, and where $\la$ denotes an affine parameter along the observer world-line. So the requirement $\tau=t$ boils down to the statement that along the  geodesic flow of the  vector field $u^\mu$  the SG coordinates $X^i$ are constant. This clearly defines the coordinate transformation in a non-perturbative way, and also shows that the geodesic observer $u_\mu= - \pa_\mu \tau$ of the GLC gauge corresponds to a static (and geodesic) observer in the synchronous gauge. It follows that the identification $t=\tau$ can always be taken for any space-time metric, and that this simple connection between  GLC  and synchronous gauge has validity far beyond the particular FLRW case or  its perturbed generalizations.

We also remark that, in GLC coordinates, the null geodesics connecting sources and observer are characterized by the simple tangent vector
 $k^{\mu} = g^{\mu \nu} \partial_{\nu} w =  g^{\mu w} = - \delta^{\mu}_{\tau} \Ups^{-1}$, which means that photons reach the observer travelling at constant $w$ and $\ti \theta^a$. This makes the calculation of the area distance and of the redshift particularly easy. Consider, for instance, a light ray emitted by a static geodesic source at the intersection between the past light-cone
of our observer, $w=w_0$, and the spatial hypersurface $\tau= \tau_s$, and received by such 
static geodesic observer at $\tau=\tau_0>\tau_s$. The associated redshift $z_s$  is then  given by \cite{GMNV}:
 \be
 \label{redshift}
 (1+z_s) = \frac{(k^{\mu} u_{\mu})_s }{(k^{\mu} u_{\mu})_o}  = \frac{(\partial^{\mu}w \pa_\mu \tau)_s }{(\partial^{\mu}w \pa_\mu \tau)_o}  = {\Ups(w_0, \tau_0, \ti \theta^a)\over \Ups(w_0, \tau_s, \ti \theta^a)} 
\, ,
 \ee
where the subscripts ``o'' and ``s'' denote, respectively, a quantity evaluated at the observer and source space-time position. The expression for the angular distance will be explicitly derived in Sect. \ref{Sec22}.

Let us finally recall that, in GLC coordinates, the covariant average of a scalar quantity $S(\tau, w, \ti \theta^a)$ over the compact two-dimensional surface $\Sg$, defined by the intersection of our past light-cone  $w=w_0$ with the spacelike  hypersurface $\tau= \tau_s$, is simply given by \cite{GMNV}:
\bea
\langle S \re_{w_0, \tau_s} &=& {\int_\Sg d^4 x \sqrt{-g}\, \da(w-w_0) \da(\tau-\tau_s) S(\tau, w, \ti \theta^a) \left| \pa_\mu \tau \pa^\mu w\right| \over 
\int_\Sg d^4 x \sqrt{-g} \,\da(w-w_0) \da(\tau-\tau_s) \left| \pa_\mu \tau \pa^\mu w\right|}
\nonumber \\
&=& {\int d^2 \ti \theta \sqrt{ \ga (w_0, \tau_s, \ti \theta^a)} \,S(w_0, \tau_s, \ti \theta^a)\over 
\int d^2 \ti \theta \sqrt{ \ga (w_0, \tau_s, \ti \theta^a)}},
\eea
where $\ga = \det \ga_{ab}$. In the case of interest for this paper, namely light-cone averages on surfaces of constant redshift $z= z_s$, one then obtains \cite{GMNV}
\beq
\langle S \re_{w_0, z_s} =  {\int d^2 \ti \theta \sqrt{ \ga (w_0, \tau (z_s, w_0, \ti \theta^a), \ti \theta^b)} \,S(w_0,  \tau (z_s, w_0, \ti \theta^a), \ti \theta^b)\over 
\int d^2 \ti \theta \sqrt{ \ga (w_0,  \tau (z_s, w_0, \ti \theta^a), \ti \theta^b)}},
\label{zaverageexact}
\eeq
where $ \tau (z_s, w_0, \ti \theta^a)$ has to be determined by solving the redshift equation (\ref{redshift}) for $\tau_s$ as a function of $w_0$, $\tau_0$, $z_s$ and $\ti \theta^a$.
This general result will now be applied to the case in which $S$ is identified with the luminosity distance $d_L$.


\subsection{Light-cone average of the luminosity distance}
\label{Sec22}
Let us first recall that the luminosity distance $d_L$ of a source at redshift $z$ is related in general to the angular distance $d_A$ of the source (as seen from the observer)  by the so-called Etherington (or reciprocity) law \cite{Et1933}:
\be
d_L = (1+z)^2 d_A\; .
\label{EqdL}
\ee
In the particular case of an unperturbed, spatially flat FLRW background, and for a source with redshift $z_s$, the distance $d_A$ is simply given by
\beq
d_A^{FLRW}(z_s)=a_s r_s= a_s(\eta_0-\eta_s),
\eeq
where $a_s= a(\eta_s)$, while $ \eta_0-\eta_s$ denotes the conformal time interval between the emission and observation of the light signal. 
For the unperturbed metric, on the other hand, we have $1+z= a_0/a(t)$, and
$d\eta= dt/a=- a_0^{-1} dz/H$, where $H= d(\ln a)/dt$. Hence:
\beq
d_L^{FLRW}(z_s) = (1+z_s) a_0 \int_{\eta_s}^{\eta_0} d\eta = (1+z_s) \int_0^{z_s} {dz\over H(z)} = {1+z_s\over H_0} \int_0^{z_s} dz \left[ \sum_n \Om_{n0} (1+z)^{3(1+w_n)}\right]^{-1/2}.
\label{standard}
\eeq
In the last equality we have used the standard (spatially flat) Friedmann equation for $H$ (see e.g. \cite{MG}), assuming that the given homogeneous model has perfect fluid sources with present fractions of the critical density $\Om_{n0}$ and barotropic parameters $w_n$. Expanding in the limit $z_s \ra 0$ we also obtain the expression
\beq
d_L^{FLRW}(z_s) \simeq 
  {1\over H_0} \left[z_s + \frac 14 \left(1 - 3 \sum_n  w_n \Om_{n0}\right) z_s^2 + O(z_s^3)\right] \equiv {1\over H_0} \left[z_s + \frac 12 (1 - q_0) z_s^2 + O(z_s^3)\right],
\label{standard1}
\eeq
which shows the well known sensitivity of the  term quadratic in $z_s$ to the composition of the cosmic fluid through the deceleration parameter $q_0$.

Let us now discuss how this well known result for $d_L$ is modified when including generic inhomogeneities. We recall, to this purpose, that in a generic metric background the angular distance $d_A$ can be computed by considering the null vector $k^\mu= dx^\mu/ d\la$ tangent to the null ray connecting source and observer (i.e. belonging to the congruence of null geodesics forming the observer's past light-cone). Here $\la$ is an affine parameter along the ray trajectory, chosen in such a way that $\la=0$ at the observer and $\la=\la_s$ at the source position. The expansion $\Theta$ of the congruence of null rays is then given by $\Theta= \nabla_\mu k^\mu$, and the corresponding angular distance $d_A$ is defined by the differential equation \cite{Maartens3, Flanagan}:
\beq {d \over d \la} \left(\ln d_A\right)= {\Theta\over 2}= {1\over 2} \nabla_\mu k^\mu.
\label{210}
\eeq

Consider now a generic metric in the GLC gauge, and the null vector  $k_\mu =\pa_\mu w$ such that $k^\mu = dx^\mu/ d \la= -\da^\mu_\tau \Ups$. We have, in this case,
\beq
\nabla_\mu k^\mu= - \Ups^{-1} \pa_\tau \left( \ln \sqrt{\ga}\right)= {d \over d \la} \left(\ln \sqrt{\ga}\right),
\eeq
where $\ga = \det \ga_{ab}$, and the integration of Eq. (\ref{210}) gives, in general,
\beq
d_A^2(\la)= c \sqrt{\ga(\la)},
\eeq
where $c$ is independent of $\la$.
In order to fix the constant $c$ we may recall the boundary conditions required to guarantee that the null rays generated by $k^\mu$ belong to the past light cones centered on the world line of our observer. In the limit $d_A \ra 0$ (or $\la \ra 0$), where $\ti \theta^a \ra \theta^a$ (see in particular next section,  Eq. (\ref{theta})),  such conditions require in particular that \cite{Maartens3}
\beq
\lim_{\la \rightarrow 0} \,{\sqrt{\ga}\over d_A^2} = \sin \theta^1_o= \sin \ti \theta^1 ,
\eeq
where, in the last equality, we have used the already mentioned fact that $\ti \theta^a$ is constant along the null geodesic. This clearly fixes $c=1/\sin \tilde{\theta}^1$, and uniquely determines the angular distance as
\be
 d_A(\la)= \ga^{1/4}(\la) \left(\sin  \tilde{\theta}^1\right)^{-1/2}.
 \label{General_Eq_dA}
\ee

Inserting this result into Eq. (\ref{zaverageexact}) we finally arrive at an exact expression for the light-cone average of the luminosity distance, as a function 
 of $z_s$, in the GLC gauge:
\beq
\langle d_L \re_{w_0, z_s} =  (1+z_s)^2 {\int d^2 \ti \theta ~ 
\ga^{1/2} (w_0,  \tau (z_s, w_0, \ti \theta^a), \ti \theta^b) 
 d_A (w_0,  \tau (z_s, w_0, \ti \theta^a), \ti \theta^b)
 \over 
\int d^2 \ti \theta ~  \ga^{1/2} (w_0,  \tau (z_s, w_0, \ti \theta^a), \ti \theta^b)}.
\label{dLaverageexact}
\eeq
This equation, being exact, can be applied in principle to any given highly inhomogeneous cosmology. Specifically, it can be applied to a  Lema\^itre-Tolman-Bondi (LTB) model \cite{LTB}
even in the case of a generic observer shifted away from the symmetry centre of the isotropic geometry. In this paper, however, we shall be mainly interested in working out the consequences of light-cone averaging for a particular model of perturbed FLRW Universe. Since we know how to describe the latter, for example, in the Newtonian (longitudinal) gauge, we need in general to connect such a coordinate system to the GLC gauge.


\section{Luminosity distance in a perturbed  FLRW geometry}
\label{Sec3}
\setcounter{equation}{0}

\subsection{First-order coordinate transformation from the Newtonian to the GLC  gauge}
\label{Sec31}

Let us consider the scalar perturbations of a conformally flat FLRW background to describe the particular 
model of inhomogeneous geometry we are interested in this paper.
Choosing in particular the so-called longitudinal or (conformally) Newtonian gauge (NG), using spherical coordinates $(r, \theta^a)= (r, \theta, \phi)$,  and going up to the first order in perturbation theory, it is well known that the model is parametrized by the following (inverse) metric tensor \cite{Muk}:
\be
g_{NG}^{\mu\nu} = a^{-2}(\eta) \,{\rm diag}\left( - 1 + 2 \Phi, 1 + 2 \Psi, (1 + 2 \Psi) \ga_0^{ab}  \right). 
\ee
Here
\beq
\ga_0^{ab}=  {\rm diag} \left( r^{-2}, r^{-2} \sin^{-2} \theta \right),
\label{gamma0}
\eeq
and  $\Phi$,  $\Psi$ are the usual gauge invariant Bardeen potentials \cite{Muk}, general functions of $\eta,r,\theta$ and $\phi$. We will assume, for simplicity, that the matter sources have vanishing (or negligible) anisotropic stress, so that $\Phi \equiv \Psi$.

For the subsequent computations -- in particular, for the 
application of the simple average prescription given previously -- 
we need to re-express this metric in GLC form, finding the transformations from the NG coordinates $y^\mu=(\eta, r, \theta, \phi)$ to the GLC coordinates
 $x^\nu=(\tau, w,\ti{\theta}^1,\ti{\theta}^2)$, and computing the reparametrized metric as
\be
g_{GLC}^{\rho\sigma}(x)=\frac{\partial x^\rho}{\partial y^\mu}
\frac{\partial x^\sigma}{\partial y^\nu} g_{NG}^{\mu\nu}(y)
\label{EqBetweenGauges}.
\ee
To this purpose we will introduce the useful (zeroth-order) light-cone variables $\eta_\pm= \eta \pm r$, such that
\beq
\eta={1\over 2} \left(\eta_+ +\eta_-\right) ~~~~~~~,~~~~~~~
r={1\over 2} \left(\eta_+ - \eta_-\right)\,,
\eeq
with corresponding partial derivatives
\beq
\pa_\eta = \pa_+ + \pa_- ~~,~~~~~ \pa_r = \pa_+ - \pa_- ~~,~~~~~\pa_\pm= {\pa \over \pa \eta_\pm}={1\over 2} \left( \pa_\eta \pm \pa_r \right).
\eeq

Using these variables we solve the three differential equations obtained from Eq. (\ref{EqBetweenGauges}) for the components $g_{GLC}^{\tau \tau} = -1$, $g_{GLC}^{ww} = 0$, $g_{GLC}^{wa} = 0$, by imposing the boundary conditions that $i)$ the transformation is non singular around the observer position at $r=0$, and $ii)$ that the two-dimensional spatial sections $r=$ const are locally parametrized at the observer's position by standard spherical coordinates, i.e. $\ti \theta^a(0)= \theta^a=(\theta, \phi)$. The sought for transformation can then be written, to first order in $\Psi$, as follows,
\bea
\label{tauwtheta}
\tau &=& \int_{\eta_{in}}^\eta d\eta' a(\eta')\left[1 + \Psi(\eta', r, \theta^a)\right] ~, \\
w &=& \eta_+ + \int_{\eta_+}^{\eta_-} dx\, \hat{\Psi}(\eta_+, x, \theta^a) ~,
\label{tauwtheta1}\\
\ti{\theta}^a &=& \theta^a + \frac{1}{2} \int_{\eta_+}^{\eta_-} dx \, \hat{\gamma}^{ab}_0 
(\eta_+, x,\theta^a) \int_{\eta_+}^x dy\, \pa_b \hat{\Psi}(\eta_+, y, \theta^a)  ~,
\label{theta}
\eea
where
$\hat{\Psi}(\eta_+,\eta_-,\theta^a)\equiv \Psi(\eta,r,\theta^a), \hat \gamma^{ab}(\eta_+,\eta_-,\theta^a)\equiv \gamma^{ab}(\eta,r,\theta^a)$
and $\eta_{in}$  represents an early enough time when the perturbation (or better the integrand) was negligible.
We can easily check that, to zeroth order in $\Psi$, we recover the homogeneous transformation (\ref{FR}) as expected.

To first order in $\Psi$ we can use again Eq. (\ref{EqBetweenGauges})
to compute the non-trivial entries of the GLC metric (\ref{GLCmetric}), and obtain:
\bea
&&
\!\!\!\!\!\!\!\!\!
\Ups = a(\eta) \left[ 1 + \hat{\Psi}(\eta_+,\eta_+, \theta^a) - \int_{\eta_+}^{\eta_-} dx \,\partial_+ \hat{\Psi}(\eta_+, x, \theta^a)\right]
+ \int_{\eta_{in}}^\eta d\eta' a(\eta')\partial_r\Psi(\eta', r, \theta^a);
\label{Ups} \\ &&
\!\!\!\!\!\!\!\!\!
U^a = \frac{1}{2}  \hat{\gamma}^{ab}_0 \int_{\eta_+}^{\eta_-} dx\, \pa_b \hat{\Psi}(\eta_+, x, \theta^a) - \frac{1}{a(\eta)}  \gamma^{ab}_0 \int_{\eta_{in}}^\eta d\eta' a(\eta') \, \pa_b
\Psi(\eta', r, \theta^a)
\nonumber   \\
&&
+\frac{1}{2} \int_{\eta_+}^{\eta_-} dx~ \partial_+
\left[\hat{\gamma}^{ab}_0(\eta_+,x,\theta^a) \int_{\eta_+}^x dy~ \pa_b\hat{\Psi}(\eta_+, y, \theta^a) \right]
-\frac{1}{2} \lim_{x\rightarrow \eta_+} \left[\hat{\gamma}^{ab}_0(\eta_+,x,\theta^a) \int_{\eta_+}^x dy~ \pa_b\hat{\Psi}(\eta_+, y, \theta^a) \right];\\
&&
\!\!\!\!\!\!\!\!\!
\gamma^{ab} = \frac{1}{a(\eta)^2}\left\{
\left[1+2 \Psi(\eta, r, \theta^a)\right]\gamma_0^{ab} 
+{1\over 2}\left[  \hat{\gamma}_0^{ac}  \int_{\eta_+}^{\eta_-} dx\,
\partial_c \left(
\hat{\gamma}_0^{bd}(\eta_+, x,\theta^a) \int_{\eta_+}^x dy\, \pa_d\hat{\Psi}(\eta_+, y, \theta^a)\right)
+ a \leftrightarrow b \right]\right\} .
\label{gammaab}
\eea
The term $\hat{\Psi}(\eta_+,\eta_+, \theta^a)$, appearing in the general equation for $\Ups$, denotes (at any $\eta$) the value of the perturbation potential evaluated at the tip of the light-cone                                                                                                                                                                                                         connecting the origin ($r=0$) to the point $y^\mu=(\eta, r, \theta^a)$ (namely, $\hat{\Psi}(\eta_+,\eta_+, \theta^a)=
\Psi(\eta+r, 0, \theta^a)$).

The above transformations can be immediately applied to obtain an explicit expression for the redshift parameter $z_s$. By inserting the result (\ref{Ups}) into  Eq. (\ref{redshift}) and considering that the source is located on the past light-cone of the observer (identified by the equation $w=w_0=\eta_0$), we obtain:
\beq
\label{OnePlusZ1}
1+z_s = \frac{a(\eta_0) }{a({\eta}_s)} \Big[1  + J(z_s, \theta^a) \Big],
\eeq
where $J =  I_+-I_r$, and where:
\bea
\label{IplusIr}
I_+ &=&\int_{\eta^s_+}^{\eta^s_-} dx \,\partial_+ \hat{\Psi}(\eta_+^s,x,\theta^a) = \Psi_s - \Psi_o - 2 \int_{\eta_s}^{\eta_0} d\eta' \, \partial_r \Psi (\eta', \eta_0-\eta', \theta^a), \\
I_r &=& \int_{\eta_{in}}^{\eta_s} d\eta' \frac{a(\eta')}{a(\eta_s)} \partial_r \Psi(\eta',\eta_0-\eta_s,\theta^a) - \int_{\eta_{in}}^{\eta_0} d\eta' \frac{a(\eta')}{a(\eta_0)} \partial_r \Psi(\eta',0,\theta^a)\,.
\label{Ir}
\eea
We have defined $\eta_\pm^s= \eta_s \pm r_s$, $\Psi_s=\Psi(\eta_s, \eta_0-\eta_s, \theta^a)$, $\Psi_o=\Psi(\eta_0, 0, \theta^a)$
and  we have used the zeroth-order light-cone condition $\eta_{+}^s =\eta_{+}^0=\eta_0$. It should be stressed, however, that while the integrals appearing in $I_r$ are evaluated at constant $r$ (namely along timelike geodesics), all the other integrals are evaluated at fixed $\eta_+$ (i.e.  along null geodesics on the observer's past light-cone). 

The contribution associated to $I_r$ can also be rewritten as
\be
\label{doppler1}
I_r =    (\vec{v}_s  - \vec{v}_o ) \cdot  \hat {n}, 
\ee
where $\hat n$ is the unit tangent vector along the null geodesic connecting source and observer, and where 
\be
\vec{v}_{s,o}=-\int_{\eta_{in}}^{\eta_{s,o}}d\eta'\frac{a(\eta')}{a(\eta_{s,0})}\vec{\nabla} \Psi(\eta', r, \theta^a)
\ee
are the  ``peculiar velocities" of source and observer associated to a  geodesic configuration perturbed up to first order in the NG gauge. 
In a realistic situation one should add to Eq.(\ref{OnePlusZ1}) similar terms taking into account a possible intrinsic (non-perturbative) motion of source and observer, unless our theoretical predictions are compared with data  already corrected for these latter Doppler contributions. Note also that our first-order expression for the redshift (\ref{OnePlusZ1}),  valid in general for any given scale factor $a(\eta)$, 
is in full agreement with the expression obtained in Eq.(38) of \cite{Bonvin}, where $z_s$ is computed for the particular case of a CDM-dominated model.


\subsection{The luminosity distance including first-order scalar perturbations}
\label{Sec32}

We now apply the above coordinate transformation  to find, in the Newtonian gauge and to first order in perturbation theory, the other relevant quantities for this paper. Let us start with the determinant $\ga$ appearing in the angular distance (\ref{General_Eq_dA}). For a source emitting light at time $\eta_s$ and radial distance $r_s$ we obtain from Eq. (\ref{gammaab}), to first order\footnote{Note that, to first order, we have $\gamma^{-1} = (\gamma_{11} \gamma_{22})^{-1}$, and that, for these diagonal matrix elements, the operator  $\pa_b$ commutes with $\gamma_0^{db}$.},
\beq
\gamma^{-1}(\la_s) \equiv \det \gamma^{ab}(\la_s) = (a_s^2r_s^2\sin\theta)^{-2} \Big[1 + 4 \Psi_s + 4 \ti{J}_2(z_s, \theta^a) \Big],
\label{det}
\eeq
 where:
 \beq
\ti{J}_2 = \frac{1}{4} \int_{\eta_+^s}^{\eta^s_-} dx\, \hat{\gamma}_0^{ab}(\eta^s_+,x,\theta^a) \int_{\eta^s_+}^x dy\, \pa_a \pa_b\hat{\Psi}(\eta^s_+,y,\theta^a) = {\frac{1}{\eta_0-\eta_s} \int_{\eta_s}^{\eta_0} d \eta' \frac {\eta' - \eta_s}{\eta_0 - \eta'} 
\Big[ \partial^2_{\theta} + (\sin\theta)^{-2}\partial^2_{\phi}\Big]  \Psi(\eta', \eta_0-\eta', \theta^a)}
\eeq
(the latter equality follows upon a simple integration by parts).
Hence, from (\ref{General_Eq_dA}):
\be
d_A(\la_s) = a_s r_s \Big[ 1 - \Psi_s  - \ti{J}_2 \Big] \left(\frac{\sin  \ti \theta^1}{\sin \theta}\right)^{-1/2}
\label{d_s2}\, .
\ee
The last factor can be easily computed, to first order, by using Eq. (\ref{theta}) and the fact 
that the $\ti{\theta}^a$ are constant along the null geodesic. It is easy to check that it amounts to a redefinition of $\ti{J}_2$, and that the above angular distance becomes
\be
d_A(\la_s)= a_s r_s \left[ 1 - \Psi_s - J_2 (z_s, \theta^a) \right]  ~,
\label{d_s2.1}
\ee
where:
\be
J_2= \frac{1}{\eta_0-\eta_s} \int_{\eta_s}^{\eta_0} d \eta \,\frac {\eta - \eta_s}{\eta_0 - \eta}  \Big[ \partial^2_{\theta} + \cot \theta\, \partial_{\theta} + (\sin \theta)^{-2} \partial^2_{\phi}\Big] \Psi(\eta', \eta0-\eta', \theta^a) \equiv \frac{1}{\eta_0-\eta_s} \int_{\eta_s}^{\eta_0} d \eta \, \frac {\eta - \eta_s}{\eta_0 - \eta} \, \Delta_2 \Psi
\label{J2_Equation}
\ee
(here $\Delta_2$ is the two-dimensional Laplacian operator on the unit two-sphere).

For the full explicit expression of the luminosity distance $d_L$ at constant redshift what we need, at this point, is the  first-order expansion of the factor $a_sr_s\equiv a(\eta_s) r_s$ appearing in Eq. (\ref{d_s2.1}). To this purpose we start from Eq. (\ref{OnePlusZ1}), considering $z_s$  as a constant parameter localising the given light source on the past light-cone   ($w=w_0$) of our observer, and we look for approximate solutions for $\eta_s= \eta_s(z_s, \theta^a)$.

Let us first define the zero-order solution ${\eta}_s^{(0)}$  through the exact relation
\be
\label{OnePlusZHom}
\frac{a({\eta}_s^{(0)})}{a_0} = \frac{1}{1+z_s},
\ee
where $a_0 \equiv a(\eta_0)$.
Expanding  (\ref{OnePlusZ1}) with respect to the parameter $\delta \eta = {\eta}_s - {\eta}_s^{(0)}$we then find:
\be
\label{OnePlusZ2}
\frac{1}{1+z_s} = \frac{a({\eta}_s^{(0)})}{a_0}  [1 + {\mathcal H}_s ~\delta \eta - J(z_s, \theta^a) ] =
\frac{1}{1+z_s} [1 + {\mathcal H}_s \, \delta \eta - J(z_s, \theta^a)]   ~, \\
\ee
where ${\mathcal H}_s= d (\ln a({\eta}_s^{(0)}))/d\eta_s^{(0)}$,
so that:
\be
 {\mathcal H}_s\, \delta \eta =  J(z_s, \theta^a)~.
 \label{320}
\ee
On the other hand, by applying Eq. (\ref{tauwtheta1}) to the light-cone $w= w_0$ at the source position, we readily obtain 
\be
w_0 = \eta^s_+  -2 \Delta \eta \Psi_{{\rm av}} = \eta_0,
\ee
where we have introduced the zero-order quantity $\Delta \eta  = \eta_0- {\eta}_s^{(0)}$, and denoted by $\Psi_{{\rm av}}$ the average value of $\Psi$ along the (unperturbed) null geodesic connecting source and observer:
\beq
\int_{\eta^s_+}^{\eta^s_-} dx \, \hat\Psi(\eta^s_+, x, \theta^a) = - 2 \int_{\eta_s}^{\eta_0} d \eta'\, \Psi (\eta', \eta_0-\eta', \theta^a) \equiv  -2 \Delta \eta \Psi_{{\rm av}}.
\label{Psi_average_Equation}
\ee
Combining this result with Eq. (\ref{320}) we can then determine, to first order, the value of the radial coordinate $r_s(z_s, \theta^a)$ corresponding to the given redshift $z_s$:
\bea
{r}_s(z_s, \theta^a) = w_0 - {\eta}_s^{(0)}(z_s) - \delta\eta  +2 \Delta \eta \Psi_{{\rm av}} 
= \Delta \eta \left[ 1 - \frac{J(z_s, \theta^a)}{{\mathcal H}_s \Delta \eta } +2  \Psi_{{\rm av}} \right].
\eea

Proceeding in the same way for $a_s$ we obtain, from Eqs. (\ref{OnePlusZ1}),
(\ref{OnePlusZHom}),
\beq
a_s(z_s, \theta^a) =  a({\eta}_s^{(0)}) \left[1+ J(z_s, \theta^a)\right],
\eeq
which, in turn, allows us to compute the value of $a_sr_s$ on the constant-$z_s$ 2-surface:
\be
[{a}_s {r}_s](z_s, \theta^a) = a({\eta}_s^{(0)}) \Delta \eta \left[ 1 +2  \Psi_{{\rm av}}  + \left(1 - \frac{1}{{\mathcal H}_s \Delta \eta}\right) J(z_s, \theta^a) \right].
\label{asrs}
\ee
The angular distance (\ref{d_s2}), for a source at redshift $z_s$,
can now be written as\footnote{Let us note that in the first order terms we can always safely identify $\eta_s$ with its unperturbed value $\eta_s^{(0)}$.}
\bea
d_A(z_s, \theta^a) =
 a({\eta}_s^{(0)}) \Delta \eta \left[1 +2  \Psi_{{\rm av}}  + \left(1 - \frac{1}{{\mathcal H}_s \Delta \eta }\right) J
 -   \Psi({\eta}_s,\eta_0-{\eta}_s,\theta^a)  - J_2\right] . 
\eea
The (first-order, non-homogeneous, non-averaged) expression of $d_L$ in our perturbed background, referred to the unperturbed value (\ref{standard}), is  finally given by:
\bea
\label{finaldL}
\frac{d_L(z_s, \theta^a)}{(1+z_s)a_0 \Delta \eta}
\equiv  {d_L(z_s, \theta^a)\over d_L^{FLRW}(z_s)} &=&
1-\Psi({\eta}_s,\eta_0-{\eta}_s,\theta^a)  +2  \Psi_{{\rm av}}   + \left(1  -  \frac{1}{{\mathcal H}_s\Delta \eta}\right) J  - J_2 .
\eea

If we apply this general result to the particular case of a CDM-dominated  Universe we find almost full agreement with the result for the luminosity distance at constant redshift computed in \cite{Bonvin}. After several manipulations, in fact, 
it turns out that our $d_L$ is equivalent to the one found in \cite{Bonvin}, modulo a term which can be written as $\vec{v}_0 \cdot \hat n$ (in the notations of Eq. (\ref{doppler1})). Such a term gives a subleading contribution to the backreaction and can be neglected with no impact on our final results.

It should be noted, however, that by expanding the above expression in the small $z_s$ limit, and comparing the result with the analogous expansion of the homogeneous distance (\ref{standard1}), we could easily introduce  a redefined value of $H_0$, say $H_0^{\rm \,ren}$, such that its inverse corresponds to the coefficient of the linear term in $z_s$ for the perturbed relation. With such a ``renormalized'' Hubble parameter we have that $H_0^{\rm \,ren} d_L$ tends smoothly to 
$H_0 d_L^{FLRW}$ for $z_s \ra 0$, and we recover full agreement with \cite{Bonvin} provided the same renormalization is applied there. Such a renormalization of $H_0$ is also suggested by (and closely related to) the first-order computation of the scalar expansion factor $\nabla_\mu u^\mu$ for the  flow wordlines  $u_\mu = \pa_\mu \tau$ of a local geodesic observer. 

At this point we could go on by taking the average of $H_0^{\rm \,ren} d_L(z_s, \theta^a)$, computing the associated backreaction, and evaluating the corrections to the standard homogeneous relation given by $H_0 d_L^{FLRW}(z_s)$. 
We have performed that exercise, but we have found that the contribution of the renormalization terms give large $z_s$-independent contributions to the variance. Furthermore, we think that renormalizing $H_0$ at $z_s=0$  is physically incorrect since, at very small-$z_s$, the backreaction is dominated  by short-scale inhomogeneities which are deeply inside the non-linear regime (where even the concept of a Hubble flow becomes inappropriate). We could instead try to renormalize $H_0$ at some small but finite $z_s$, e.g. at a redshift corresponding to the closest used supernovae (say $z_s \sim 0.015$, see e.g. \cite{SN}), but then the results (although much better behaved) would depend on the choice of the particular ``renormalization point". Thus, it seems best to consider just the full ``unrenormalized" expression (\ref{finaldL})  in a limited region of $z_s$ where one can trust the approximations  made, and use that expression for a phenomenological parametrization of the backreaction effects which could possibly include a redefinition of $H_0$.


\section{Combining space-time and {\em ensemble} averages}
\label{Sec4}
\setcounter{equation}{0}

In the following sections the inhomogeneous deviations from the standard FLRW quantities are sourced by a stochastic background of primordial perturbations,  satisfying $\overline \Psi=0$, $\overline{\Psi^2} \not=0$, where the bar denotes statistical (or {\em  ensemble}) average 
(see Sec. \ref{Sec5}). Hence,  if we limit ourselves to a first-order computation of $d_L$,  we immediately obtain  
$\overline{d_L}= d_L^{FLRW}$. Non-trivial effects can only be obtained from quadratic and higher-order perturbative corrections, or from the spectrum of the two-point correlation function $\overline{d_L(z, \theta^a) d_L(z', \theta^{\prime a})}$ (discussed in detail in \cite{Bonvin}). 
 
In this paper we will consider the  {\em ensemble} average not of $d_L$ but of $\langle d_L \re$, where the angular brackets refer to the light-cone average defined in Eq. (\ref{dLaverageexact}) (see e.g. \cite{Li:2007ny,precision,CU} for previous attempts of combining {\em ensemble} average with averages over spacelike hypersurfaces). We will see that the  light-cone average automatically induces quadratic (and higher-order) backreaction terms, due to the inhomogeneities present both in $d_L$ and in the covariant integration measure, and since  the {\em ensemble} average of such terms is non-vanishing we obtain, in general, $\overline{\langle d_L \re} \not= d_L^{FLRW}$. 

We will start this section  with some general considerations on how to combine space-time and {\em ensemble} averaging, and how to isolate those terms in $\overline{\langle d_L \re}$ that we may genuinely call ``backreaction" effects, i.e. effects on averaged quantities due to inhomogeneities. We shall also discuss how to estimate the variance around mean values due to such fluctuations\footnote{The importance of the cosmic variance for a precise measurement of the cosmological parameters, taking into account  backreaction effects from
averaging on domains embedded in a spatial hypersurface (according to \cite{1}), has been recently pointed out also in \cite{WS}.}. Many of these considerations can be certainly found elsewhere, but are nonetheless  presented here  for the sake of being self-contained.

Let us consider a typical  average over the compact surface $\Sg$ (topologically equivalent to a two-sphere) embedded on the past light-cone $w=w_0$ at constant $z_s$. We simply denote such an average by:
\beq
\label{AvS}
\langle S  \rangle_\Sg =  \frac{\int_\Sg d^2 \mu \,S}{ \int_\Sg d^2 \mu} ~,
\eeq
where $d^2 \mu$ is the appropriate measure provided by  our gauge invariant prescription (see Eq. (\ref{zaverageexact})) and $S$ is the (possibly non local) scalar observable ($d_L$ in our case).
We can conveniently extract,  from both $d^2 \mu$ and from  $S$, a zeroth-order homogeneous contribution by defining:
\bea
\label{exp}
d^2 \mu =  (d^2 \mu)^{(0)}(1 + \mu), ~~~~~~~~~~  S = S^{(0)} (1 + \sigma), 
\eea
and use the possibility of rescaling both integrals in Eq. (\ref{AvS}) by the same constant, in order to normalize $\int (d^2 \mu)^{(0)} = 1$.
We then easily get:
\bea
\label{AvS1}
\left\langle \frac{S}{S^{(0)}}  \right\rangle =  \frac{\int (d^2 \mu)^{(0)}(1+ \mu)(1+\sigma) }{\int (d^2 \mu)^{(0)}(1+ \mu)} = 
{\langle (1+ \mu)(1+\sigma)\rangle_0  \over 1+ \langle \mu \rangle_0} ~,
\eea
where we have dropped, for simplicity, the subscript of the averaging region $\Sg$, and where we have defined by $\langle \dots \rangle_0$ averages with respect to the unperturbed measure $(d^2 \mu)^{(0)}$ (we shall drop the subscript $0$ hereafter).

Let us now perform the {\em ensemble} average, denoted by an overbar, paying attention to the fact that {\em ensemble} averages do {\it not} factorize, i.e. $\overline{AB} \ne \overline{A} ~\overline{B}$. A simple calculation leads to:
\bea
\label{EnsAv}
\overline{\langle S/S^{(0)}  \rangle}  =  1+ \overline{(\langle \sigma\rangle+ \langle \mu \sigma\rangle)  (1+ \langle \mu \rangle)^{-1}} ~.
\eea
This last equation is supposedly exact but, as such, pretty useless. It becomes an interesting equation, though, if we can expand the quantities $\mu$ and $\sigma$ in a perturbative series:
\bea
\mu = \sum_i \mu_i, ~~~~~ ~~~~~~ \sigma = \sum_i \sigma_i, 
\eea
and we further assume that the first order quantities $\mu_1, \sigma_1$ have vanishing {\em ensemble} averages (as it is the case for typical cosmological perturbations coming from inflation).
In that case we can easily expand the result and obtain, for instance:
\beq
\label{EnsAvExp}
\overline{\langle S/S^{(0)}  \rangle}  = 1 +  \overline{\langle \sigma_2 \rangle} + \rm{IBR}_2 + \overline{\langle \sigma_3 \rangle} + \rm{IBR}_3 + \dots 
\eeq
where
\bea
 \rm{IBR}_2 &=&  \overline{\langle \mu_1 \sigma_1\rangle} - \overline{\langle \mu_1\rangle  \langle \sigma_1\rangle},  \label{ibr2}\\
 \rm{IBR}_3 &=&  \overline{\langle \mu_2 \sigma_1\rangle} - \overline{\langle \mu_2\rangle  \langle \sigma_1\rangle} +
\overline{\langle \mu_1 \sigma_2\rangle} - \overline{\langle \mu_1\rangle  \langle \sigma_2\rangle} 
 - \overline{\langle \mu_1\rangle  \langle \mu_1 \sigma_1\rangle} + \overline{\langle \mu_1\rangle \langle \mu_1\rangle \langle \sigma_1\rangle},
\eea
and where we have used again the non-factorization property, i.e. $ \overline{\langle \mu_1\rangle  \langle \sigma_1\rangle} \neq \overline{\langle \mu_1\rangle}~  \overline{\langle \sg_1\rangle}$, and so on. 
We see that the result contains, to a given order, both terms that depend on expanding $S$ to that order (but {\it not} on the precise averaging prescription), and ``induced backreaction'' (IBR) terms  that  depend on correlations between the fluctuations of $S$ and those in the measure. These latter terms only depend on lower-order perturbations of $S$ and the measure separately. In particular, our first-order calculation provides the full second-order IBR effect that comes from the above interplay of $\mu_1$ and $\sg_1$, although the full second-order result needs also the harder computation of $\sigma_2$ (but {\em not} of $\mu_2$). Note also that, whenever the fluctuations of $S$ and $d^2 \mu$ are uncorrelated, all IBR effects drop out. We will see below how to apply the above general reasoning to the particular case of $d_L$.

Let us now discuss instead the issue of the variance, i.e. of how broad is the distribution of values for $S/S^{(0)} $ around its mean value $\overline{\langle S/S^{(0)}  \rangle}$. This dispersion is due to both the fluctuation on the averaging surface and to those due to {\em ensemble} fluctuations. Let us thus define:
 \begin{equation}
 \label{vargen}
{\rm Var}[S/S^{(0)}] \equiv \overline{\lla \left(S/S^{(0)} - \overline{\langle S/S^{(0)}\rangle} \right)^2 \rra}  = \overline{\lla (S/S^{(0)})^2 \rra}  - \left(\overline{\lla S/S^{(0)} \rra}\right)^2.
\end{equation}
Inserting the definition (\ref{exp}) we get, after a little algebra:
  \begin{equation}
  \label{vargenexp}
{\rm Var}[S/S^{(0)}] =  \overline{\langle \sigma^2 (1 + \mu) \rangle(\langle 1 + \mu \rangle)^{-1}}  - \left(\overline{\langle \sigma (1 + \mu) \rangle (1+ \langle  \mu \rangle)^{-1}} \right)^2. 
\end{equation}
If we now make the same assumptions as before on expanding $\mu$ and $\sigma$, we find that the second term
in (\ref{vargenexp}) is at least of fourth order.  
Therefore, for the leading term in the variance we find the amazingly simple result (see also \cite{WS}):
\begin{equation}
  \label{vargen2nd}
{\rm Var}[S/S^{(0)}] =  \overline{\langle \sigma_1^2  \rangle}.
\end{equation}
As in the case of IBR$_2$ we only need to know the first-order perturbation, but this time  the effect is completely independent of the averaging measure. As we shall see, the dispersion (which is actually the square root of the variance) turns out to be larger than the averaging corrections due to IBR$_2$.

We may note, at this point, that we could have  also considered the dispersion of the angular average $\langle S/S^{(0)}\rangle$ due to the stochastic fluctuations, namely:
 \begin{equation}
 \label{vargen1}
{\rm Var'}[S/S^{(0)}] \equiv \overline{\left(\langle S/S^{(0)}\rangle- \overline{\langle S/S^{(0)}\rangle} \right)^2}  = 
\overline{\left(\lla S/S^{(0)}\rra\right)^2}  - \left(\overline{\lla S/S^{(0)} \rra}\right)^2\,,
\end{equation}
which, after calculations similar to the ones above, gives
\be 
\label{varp}
{\rm Var'}[S/S^{(0)}]   = \overline{\langle \sigma_1 \rangle^2}\,.
\ee
Such a quantity is much smaller than the previous one, as can be inferred from Section \ref{Sec6},  indicating that the main reason for the dispersion lies in the angular scatter of the data rather than in their stochastic distribution due to the {\em ensemble}.

Let us finally identify the quantities appearing in $\rm{IBR}_2$ for our particular case. To this purpose let us stress that, to describe the real impact of  inhomogeneties on the observational data,  we should consider the  scalar $S$ corresponding to the true observed quantity. In the case of the supernovae data \cite{nobel} this quantity should be the received flux of radiation, which is proportional to $\sim d_L^{-2}$. On the other hand, if we  consider the average of  $S=d_L$ instead of the average of the flux, the difference is only of 
second order in our perturbative expansion~\footnote{Since the  flux is proportional to $d_L^{-2}(z_s, \theta^a)$   the distance modulus  (see Sec. \ref{Sec6})
should be computed  as a function of $(\overline{\langle d_L^{-2} \rangle})^{-1/2}$ rather than of   $\overline{\langle d_L \rangle}$. Up to the second  order the difference between the two  is shown in the following expressions:
\be
(\overline{\langle d_L^{-2} \rangle})^{-1/2}=d_L^{FLRW} \left(1+\overline{\langle \sigma_2 \rangle}+IBR_2-\frac{3}{2}\overline{\langle \sigma_1^2 \rangle}\right) ~;~~ ~~~~~~~~~~~
\overline{\langle d_L \rangle}=d_L^{FLRW} \left(1+\overline{\langle \sigma_2 \rangle}+IBR_2\right). \nonumber
\ee}. 
Since in this paper we will not evaluate such genuine second-order contribution to the average we can limit ourselves,  for the sake of simplicity,  to consider hereafter $S=d_L(z_s, \theta^a)$.

In such a case, from Eq. (\ref{finaldL}) we immediately  find
\beq 
\label{sigma1}
\sigma_1 = A_1+A_2+A_3 +A_4+A_5,
\eeq
where we have defined:
\bea
\label{defA}
A_1 &=& -\Psi_s ~;~~~~  A_2 =  2  \Psi_{{\rm av}} ~;~~~~  A_3 =   \left(1  -  \frac{1}{{\mathcal H}_s\Delta \eta}\right) I_+ ~;~~~~   A_4 = -\left(1  -  \frac{1}{{\mathcal H}_s\Delta \eta}\right) I_r ~; ~~~~ A_5= - J_2.
\eea
On the other hand, in our particular case, the general measure $d^2\mu$  will be  given by
Eq. (\ref{zaverageexact}):
\be
d^2\mu=d^2 \ti{\theta} \sqrt{\gamma(w_0,\tau(z_s,w_0,\tilde{\theta}^a),\tilde{\theta}^a)}\,.
\ee 
We need  to transform the integral in $d^2 \ti{\theta}$ (over the ``$2$-sphere" $\Sg$)  to the standard polar coordinates  $d^2 \theta= d\theta d\phi$ of the Newtonian  gauge. To first order, it is easy to check (by taking into account the Jacobian determinant of the transformation $\theta^a \ra \ti \theta^a$, see Eq. (\ref{theta})) that:
\be
\int d^2 \ti{\theta} \sqrt{\gamma} = \int d\phi d\theta \sin \theta \left([a_s r_s](z_s, \theta^a)\right)^2 (1- 2 \Psi_s) ~.
\ee
The normalized unperturbed measure is then given by  $(d^2 \mu)^{(0)}= d^2 \Om/ 4 \pi$, where $d^2\Om= d \phi d\theta \sin \theta$.
Proceeding as in the previous section, and using in particular Eq. (\ref{asrs}),  we also obtain:
\be
\label{mu1}
\mu_1= -2 \Psi_s  + 4 \Psi_{{\rm av}} + 2 \left(1  -  \frac{1}{{\mathcal H}_s\Delta \eta}\right) J(z_s, \theta^a) = 2( A_1 + A_2 + A_3 + A_4) \,.
\ee

The second-order induced backreaction IBR$_2$, and the variance of $d_L/d_L^{\rm FLRW}$,  can now be obtained by inserting the results (\ref{sigma1}), (\ref{mu1}) into Eqs. (\ref{EnsAvExp}) and (\ref{vargen2nd}), respectively. Their explicit evaluation, first analytic and then numerical, will be presented in the next two sections.


\section{Induced backreaction and dispersion: analytic considerations}
\label{Sec5}
\setcounter{equation}{0}

This section will be devoted to a systematic evaluation of the various terms contributing to the induced backreaction IBR$_2$ defined in the previous section, as well as to the variance associated to the (angle and {\em ensemble})-averaged value of $d_L(z_s, \theta^a)$.

Let us start by noting that the simplest way to implement the {\em ensemble} average of our stochastic background of scalar perturbations $\Psi$ is to consider their Fourier decomposition in the form:
\be
\Psi(\eta, \vec{x}) = \frac{1}{(2 \pi)^{3/2}} \int d^3 k \, \e^{i\vec{k}\cdot \vec{x}} \Psi_k(\eta) E(\vec{k}) \,,
\label{PsiFourier}
\ee
where -- assuming that the fluctuations are statistically homogeneous -- $E$  is a unit random variable satisfying $E^*(\vec{k})=E(-\vec{k})$ as well as  the {\em ensemble}-average condition:
\be
\overline{E(\vec{k}_1) E(\vec{k}_2)}=\delta(\vec{k}_1+\vec{k}_2).
\label{PropRandomVar}
\ee

According to the results of Sec. \ref{Sec4}, all corrections we need to compute are bilinear terms in the potential $\Psi$ always occurring in the combination $\overline{\langle A_i A_j \rangle}$ or  $\overline{\langle A_i \rangle \langle A_j \rangle} $, where  the quantities $A_i$, $A_j$ are defined in Eq. (\ref{defA}). Let us illustrate a typical computation with one of the simplest contributions associated to $A_1= -\Psi_s$. Following the notations of Sect. \ref{Sec4} we will denote with the angular brackets the integration over the two-surface $\Sg$ embedded on the 
light-cone  (at $w=w_0$, $z=z_s$). The measure of integration is the unperturbed normalized one, i.e. 
$d^2 \Om/ 4 \pi$. We then obtain:
\bea
\overline {\lla \Psi_s \Psi_s \rra} &=& \int \frac{d^3 k ~d^3 k'}{(2 \pi)^{3}} \overline{E(\vec{k}) E(\vec{k'})} \int \frac{d^2 \Omega}{4 \pi} \left[ \Psi_k(\eta_s) e^{i r \vec{k}\cdot \hat{{x}}}\right]_{r=\eta_0 - \eta_s} \cdot \left[ \Psi_{k'}(\eta_s) e^{i r  \vec{k'}\cdot\hat{{x}}} \right]_{r=\eta_0 - \eta_s} \nonumber \\
&=&  \int \frac{d^3 k}{(2 \pi)^{3}} ~ |\Psi_k(\eta_s)|^2 \int_{-1}^{1} \frac{d(\cos\theta)}{2} \left[ e^{i k \Delta \eta \cos\theta} \right] \cdot \left[ e^{-i k \Delta \eta \cos\theta} \right] \nonumber \\
&=&  \int \frac{d^3 k}{(2 \pi)^{3}}  ~ |\Psi_k(\eta_s)|^2 = \int_0^{\infty} \frac{d k}{k}  ~ P_{\Psi}(k,\eta_s),
\\
\overline {\lla \Psi_s \rra \lla \Psi_s \rra} &=& \int \frac{d^3 k ~d^3 k'}{(2 \pi)^{3}} \overline{E(\vec{k}) E(\vec{k'})} \left[ \int \frac{d^2 \Omega}{4 \pi} \Psi_k(\eta_s) e^{i r \vec{k}\cdot \hat{{x}}} \right]_{r=\eta_0 - \eta_s} \cdot \left[ \int \frac{d^2 \Omega'}{4 \pi} \Psi_{k'}(\eta_s) e^{i r  \vec{k'}\cdot\hat{{x}}'} \right]_{r=\eta_0 - \eta_s} \nonumber \\
&=&  \int \frac{d^3 k}{(2 \pi)^{3}} ~ |\Psi_k(\eta_s)|^2 \left[ \int_{-1}^{1} \frac{d(\cos\theta)}{2} e^{i k \Delta \eta \cos\theta} \right] \cdot \left[ \int_{-1}^{1} \frac{d(\cos\theta')}{2} 
e^{-i k \Delta \eta \cos\theta'} \right] \nonumber \\
&=&  \int \frac{d^3 k}{(2 \pi)^{3}}  ~ |\Psi_k(\eta_s)|^2 \left( \frac{\sin(k \Delta\eta)}{k \Delta \eta} \right)^2 = 
 \int_0^{\infty} \frac{d k}{k}  ~ P_{\Psi}(k,\eta_s) \left( \frac{\sin(k \Delta\eta)}{k \Delta \eta} \right)^2,
\eea
where in the second line of both terms we made use of isotropy (i.e. $\Psi_k$ only  dependent on $k = |\vec k|$), and defined $\theta$ and $\theta'$ as the angles between $\vec{k}$ and $\vec{x} \equiv r \hat{{x}}$ and between $\vec{k}'$ and $\vec{x}' \equiv r \hat{{x}}'$. We have also introduced the (so-called dimensionless) power spectrum of $\Psi$, defined in general by:
\be
P_\Psi (k, \eta)  \equiv \frac{k^3}{2 \pi^2}  |\Psi_k(\eta)|^2 .
\ee
More complicated examples, that contain almost all the subtleties of these computations, will be presented in Appendix A.

In general we have many contributions like the above ones, appearing in both  the induced backreaction and the variance, and generated by all the $A_i$ terms of Eq. (\ref{defA}). In this paper we will consider the particularly simple  case of a CDM-dominated background geometry, with a time-independent spectral distribution of sub-horizon scalar  perturbations, $\pa_\eta \Psi_k=0$. 
In such a case all the relevant contributions  can be parameterized in the form
\bea
\overline{\langle A_iA_j \rangle} &=& \int_0^{\infty} \frac{d k}{k}  ~ P_{\Psi}(k)  {\mathcal C}_{ij}(k,\eta_0,\eta_s),  
\label{aiaj}
\\  
\overline{\langle A_i \rangle \langle  A_j \rangle} & =& \int_0^{\infty} \frac{d k}{k}  ~ P_{\Psi}(k) {\mathcal C}_{i}(k,\eta_0,\eta_s)~ {\mathcal C}_{j} (k,\eta_0,\eta_s)
\label{ai}
\eea
(where ${\mathcal C}_{ij}, {\cal C}_i$ are constant spectral coefficients), valid for any given model of (time-independent) scalar perturbation spectrum. With such parametrization, the leading-order induced backreaction,  called $\rm{IBR}_2$ in the expansion \rref{EnsAvExp} of $\overline{\langle d_L  \rangle}/ d_L^{FLRW}$, can be written as
\begin{equation}
{\rm IBR}_2 =  2 \sum_{i=1}^4 \sum_{j=1}^5  \Big[ \overline{\langle A_iA_j \rangle}  -\overline{\langle A_i \rangle \langle  A_j \rangle}  \Big] = \int_0^{\infty} \frac{d k}{k}  ~ P_{\Psi}(k)\, 2 \sum_{i=1}^4 \sum_{j=1}^5  \Big[ {\mathcal C}_{ij}(k,\eta_0,\eta_s)  - {\mathcal C}_{i}(k,\eta_0,\eta_s)~ {\mathcal C}_{j} (k,\eta_0,\eta_s) \Big]\,.
\label{BR2backreaction}
\end{equation}
The dispersion (from Eq.(\ref{vargen2nd})) takes instead the form:
\begin{equation}
\label{BRVarTh}
\left({\rm Var}\left[\frac{d_L}{d_L^{FLRW}}\right]\right)^{1/2} = \sqrt{\overline{\lla  \sigma_1^2 \rra}} = \left[ \sum_{i=1}^5 \sum_{j=1}^5 \overline{\langle A_iA_j \rangle}\right]^{1/2} =
\left[\int_0^{\infty} \frac{d k}{k}  ~ P_{\Psi}(k) \sum_{i=1}^5 \sum_{j=1}^5  {\mathcal C}_{ij}(k,\eta_0,\eta_s)\right]^{1/2}.
\end{equation}

We have analytically computed all the required coefficients ${\mathcal C}_{ij}, {\cal C}_i$, and their final values are given in Table I for  ${\mathcal C}_{ij}(k,\eta_0,\eta_s)$, and in Table II for ${\mathcal C}_{i}(k,\eta_0,\eta_s)$. In Table I we also show  the small-$k$ limit ($k \Delta \eta \ll 1$) of ${\mathcal C}_{ij}$ and of the products ${\mathcal C}_{i} {\mathcal C}_{j}$. For notational convenience we have introduced in the tables the dimensionless variable $l= k \Da \eta$, and we have used the definition:
\beq
{\rm SinInt}(l)= \int_0^l {dx\over x} \sin x.
\eeq

\begin{table}[ht!]
\caption[]{\label{Tab1} The spectral coefficients ${\mathcal C}_{ij}(k,\eta_0,\eta_s)$ for the  $\overline{\lla A_i A_j \rra}$ terms defined by Eq. (\ref{aiaj}). We also give the $k \Delta \eta \ll 1$ limit (up to  leading order in $k \Delta \eta$) of ${\mathcal C}_{ij}$ and of the products ${\mathcal C}_{i} {\mathcal C}_{j}$ for the coefficients defined in Table II.}
\vskip 0.3 cm
\begin{tabular}{|c|c|c|c|}
\hline
$\overline{\lla A_i A_j \rra}$ & ${\mathcal C}_{ij}(k,\eta_0,\eta_s)$ & ${\mathcal C}_{ij}$ for $k \Delta \eta \ll 1$ & ${\mathcal C}_i~{\mathcal C}_j$ for $k \Delta \eta \ll 1$\\
\hline\hline
$\overline{\lla A_1 A_1 \rra}$ & 1 & 1 & $1-\frac{l^2}{3}$ \\ \hline
$\overline{\lla A_1 A_2 \rra}$ & $-\frac{2}{l} {\rm SinInt}(l)$ & $-2+\frac{l^2}{9}$ & $-2 +\frac{4}{9} l^2$\\ \hline
$\overline{\lla A_1 A_3 \rra}$ & $\left(1 - \frac{1}{{\mathcal H}_s\Delta \eta}\right)  \left[1 - \frac{\sin(l)}{l} \right]$ & $\left(1 - \frac{1}{{\mathcal H}_s\Delta \eta}\right)\frac{l^2}{6}$ & $-\left(1 - \frac{1}{{\mathcal H}_s\Delta \eta}\right)\frac{l^2}{6}$ \\ \hline
$\overline{\lla A_1 A_4 \rra}$ & $\left(1 - \frac{1}{{\mathcal H}_s\Delta \eta}\right) 
\frac{f_0}{\Delta \eta}[\cos l - \frac{\sin(l)}{l}]$ & 
$-\frac{f_0}{\Delta \eta} \left(1 - \frac{1}{{\mathcal H}_s\Delta \eta}\right) \frac{l^2}{3}$ & 
$-\frac{f_s}{\Delta \eta} \left(1 - \frac{1}{{\mathcal H}_s\Delta \eta}\right) \frac{l^2}{3}$ \\ \hline
$\overline{\lla A_1 A_5 \rra}$ & $- 2 \left[1 - \frac{\sin(l)}{l} \right]$ & 
$-\frac{l^2}{3}$ & 
0 \\ \hline
$\overline{\lla A_2 A_2 \rra}$ & $\frac{8}{l^2} \left[ -1 + \cos l + l {\rm SinInt}(l) \right]$ & 
$4-\frac{l^2}{9}$ & 
$4-\frac{4}{9} l^2$ \\ \hline
$\overline{\lla A_2 A_3 \rra}$ & 0 & 
0 & 
$\left(1 - \frac{1}{{\mathcal H}_s\Delta \eta}\right) \frac{l^2}{3}$\\ \hline
$\overline{\lla A_2 A_4 \rra}$ & $2 \left(1 - \frac{1}{{\mathcal H}_s\Delta \eta}\right)  \frac{f_0 + f_s}{\Delta \eta}[1-\frac{\sin(l)}{l}]$ & 
$\frac{f_0+f_s}{\Delta \eta} \left(1 - \frac{1}{{\mathcal H}_s\Delta \eta}\right) \frac{l^2}{3}$ & 
$\frac{f_s}{\Delta \eta} \left(1 - \frac{1}{{\mathcal H}_s\Delta \eta}\right) \frac{2}{3}l^2$ \\ \hline
$\overline{\lla A_2 A_5 \rra}$ & $\frac{2}{3l^2} \left[-4 + (4 + l^2)\cos l + l \sin l + l^3 {\rm SinInt}(l) \right]$ & 
$\frac{l^2}{3}$ & 
0 \\ \hline
$\overline{\lla A_3 A_3 \rra}$ & $2 \left(1 - \frac{1}{{\mathcal H}_s\Delta \eta}\right)^2 \left[ 1 - \frac{\sin(l)}{l} \right]$ & 
$\left(1 - \frac{1}{{\mathcal H}_s\Delta \eta}\right)^2 \frac{l^2}{3} $ & 
$\left(1 - \frac{1}{{\mathcal H}_s\Delta \eta}\right)^2 \frac{l^4}{36}$ \\ \hline
$\overline{\lla A_3 A_4 \rra}$ & $\left(1 - \frac{1}{{\mathcal H}_s\Delta \eta}\right)^2  \frac{f_0 - f_s}{\Delta \eta} \left[ \cos l - \frac{\sin(l)}{l} \right]$ & 
$-\frac{f_0-f_s}{\Delta \eta} \left(1 - \frac{1}{{\mathcal H}_s\Delta \eta}\right)^2 \frac{l^2}{3} $ & 
$\frac{f_s}{\Delta \eta} \left(1 - \frac{1}{{\mathcal H}_s\Delta \eta}\right)^2 \frac{l^4}{18}$ \\ 
\hline
$\overline{\lla A_3 A_5 \rra}$ & $-2 \left(1 - \frac{1}{{\mathcal H}_s\Delta \eta}\right)  \left[1 - \frac{\sin(l)}{l} \right]$ & 
$- \left(1 - \frac{1}{{\mathcal H}_s\Delta \eta}\right) \frac{l^2}{3}$ & 
0 \\ \hline
$\overline{\lla A_4 A_4 \rra}$ & $\left(1 - \frac{1}{{\mathcal H}_s\Delta \eta}\right)^2  \left[ \frac{f_0^2 + f_s^2}{\Delta \eta^2} \frac{l^2}{3} - \frac{2 f_0 f_s}{\Delta \eta^2} \left( 2 \cos l + (-2 + l^2) \frac{\sin l}{l} \right) \right]$ & 
$\left(\frac{f_0-f_s}{\Delta \eta}\right)^2 \left(1 - \frac{1}{{\mathcal H}_s\Delta \eta}\right)^2 \frac{l^2}{3}$ & 
$\left(\frac{f_s}{\Delta \eta}\right)^2 \left(1 - \frac{1}{{\mathcal H}_s\Delta \eta}\right)^2 \frac{l^4}{9}$ \\ \hline
$\overline{\lla A_4 A_5 \rra}$ & $\left(1 - \frac{1}{{\mathcal H}_s\Delta \eta}\right) \left[\frac{f_0 + 3 f_s}{\Delta \eta} \cos l + \frac{f_0 - f_s}{\Delta \eta} \frac{\sin l}{l} + \frac{(f_0 + f_s) (-2 + l {\rm SinInt}(l))}{\Delta \eta} \right]$ & 
$\frac{f_0-f_s}{\Delta \eta} \left(1 - \frac{1}{{\mathcal H}_s\Delta \eta}\right) \frac{l^2}{3}$ & 
0 \\ \hline
$\overline{\lla A_5 A_5 \rra}$ & $\frac{1}{15 l^2} \left[-24 + 20 l^2 + (24 - 2 l^2 + l^4) \cos(l) \right.$ & 
$\frac{l^2}{3}$ & 
0 \\
  & $\left. + l (-6 + l^2) \sin(l) + l^5 {\rm SinInt}(l) \right]$ &  &  \\ \hline
\end{tabular}
\end{table}

\begin{table}[ht!]
\centering
\caption[]{\label{Tab2} The spectral coefficients ${\mathcal C}_i$ for the  $\overline{\lla A_i \rra \lla A_j \rra}$ terms defined by Eq. (\ref{ai}).}
\vskip 0.3 cm
\begin{tabular}{|c|c|}
\hline
$A_i$ & ${\mathcal C}_i(k,\eta_0,\eta_s)$ \\
\hline\hline
$A_1$ & $\frac{\sin l}{l}$ \\ \hline
$A_2$ & $-\frac{2}{l} {\rm SinInt}(l)$ \\ \hline
$A_3$ & $-\left(1  -  \frac{1}{{\mathcal H}_s\Delta \eta}\right) \left(1 - \frac{\sin l}{l}\right)$ \\ \hline
$A_4$ & $ \left(1  -  \frac{1}{{\mathcal H}_s\Delta \eta}\right) \frac{f_s}{\Delta \eta} \left( \cos l - \frac{\sin l}{l} \right)$ \\ \hline
$A_5$ & 0 \\ \hline
\end{tabular}
\end{table}

It should be noted that the computation of some of the above terms requires the explicit expression of the scale factor $a(\eta)$. In those cases we have assumed  a dust-dominated scale factor with $a(\eta) = a(\eta_0)(\eta/\eta_0)^2$, and we have defined
\beq
f_{0,s} \equiv \int_{\eta_{in}}^{\eta_{0,s}} d\eta \frac{a(\eta)}{a(\eta_{0,s})} = \frac{\eta_{0,s}^3 - \eta_{in}^3}{3\eta_{0,s}^2} ~ \simeq \frac13 \eta_{0,s}
\eeq
(recall that $\eta_{in}$ satisfies, by definition, $\eta_{in} \ll \eta_{0,s}$). 
Using such a CDM-dominated model also imposes the following (zeroth-order) relation between $z_s$ and $\Delta \eta$:
\be
\Delta \eta = \frac{2}{a_0H_0} \left[1-(1+z_s)^{-1/2}\right] ~.
\ee
This relation will be used in section \ref{Sec6} when performing the numerical integration over $k$ as a function of  $z_s$. 

In order to proceed with some qualitative and quantitative considerations it is important, at this point,  to specify the  properties of the power spectrum $P_{\Psi}(k)$. Limiting ourselves to sub-horizon perturbations, and considering the standard CDM model, we can simply obtain  
$\Psi_k$ by applying an appropriate, time-independent transfer function to the primordial (inflationary) spectral distribution (see e.g. \cite{DurCos}). The power spectrum of the Bardeen potential is then given by
\be
\label{PsiP}
P_\Psi (k)  = \left(\frac{3}{5}\right)^2
\Delta_{\cal R}^2 T^2(k) ~ , ~~~~~~~~~~~~~ \Delta_{\cal R}^2=A \left(\frac{k}{k_0}\right)^{n_s-1} ~,
\ee
where $T(k)$ is a constant transfer function which takes into account the sub-horizon evolution of the modes re-entering during the radiation-dominated era, and  $\Delta_{\cal R}^2$ is the primordial power spectrum of  curvature perturbations, amplified by inflation, outside the horizon. The typical parameters of such a spectrum, namely the amplitude $A$,  the spectral index $n_s$ and the scale $k_0$, are determined by the results of recent WMAP observations  \cite{WMAP7}. In our computations we will use, in particular, the following approximate values:
\beq
 A=2.45 \times 10^{-9} ~, ~~~~~~~~  n_s=0.96 ~, ~~~~~~~~ k_0/a_0=0.002 \,{\rm Mpc}^{-1} ~.
 \eeq

Finally, since we are mainly interested in the overall magnitude of the transfer function, it will be enough for our needs  to approximate $T(k)$ with the effective shape of the transfer function for density perturbations\footnote{The relation between density perturbations and metric perturbations is of course fully under control in the  linear perturbative regime, but could be unreliable in the high-$k$ non-linear regime.}    with no baryon oscillations, namely $T(k)=T_0(k)$, where \cite{Eisenstein:1997ik}:
\beq
\label{EH97}
T_0(q) = \frac{L_0}{L_0+q^2C_0(q)}  ~, ~~~~ L_0(q) = \ln(2e+1.8q) ~, 
~~~~ C_0(q) = 14.2+\frac{731}{1+62.5q} ~, ~~~~q  = \frac{k}{13.41 k_{\rm eq}} ~,
\eeq
and where $k_{\rm eq}$ is the scale corresponding to matter-radiation equality. We can easily see that the above transfer function goes to 1 for $k \ll k_{\rm eq}$,  while it falls like  $k^{-2}\log k$ for $k \gg k_{\rm eq}$.
In the next section we will numerically evaluate the  coefficients of Tables I and II, and we will discuss the corresponding backreaction  effects on $\overline{\langle d_L \re}$, together with its dispersion, for the scalar perturbation spectrum described by Eqs. (\ref{PsiP}) and (\ref{EH97}).


\section{Induced backreaction and dispersion: numerical results and discussion}
\label{Sec6}
\setcounter{equation}{0}

In the following computations we will set $a_0 =1$, $\Omega_m =1$, and we will use $h\equiv H_0/(100 \,{\rm km \,s}^{-1} {\rm Mpc}^{-1})=0.7$. In that case we obtain \cite{Eisenstein:1997ik} 
$k_{\rm eq}\simeq 0.036 \,{\rm Mpc}^{-1}$, 
and we can more precisely define the asymptotic regimes of our transfer function as $T_0 \simeq 1$ for $k \laq 10^{-3}\, {\rm Mpc}^{-1}$, and $T_0 \sim k^{-2}\log k$ for  $k \gaq 2.5\, {\rm Mpc}^{-1}$.
This second scale is already deep inside the so-called non-linear regime, roughly corresponding to $k \gaq 0.1\, h \, {\rm Mpc}^{-1}$ (see e.g. \cite{scoc}) . 
 
Let us  start by comparing the behaviour of the power spectrum with the behaviour of the coefficients in Table \ref{Tab1} and \ref{Tab2}. As shown in Table \ref{Tab1}, all the coefficients ${\mathcal C}_{ij}$ and ${\mathcal C}_{i} {\mathcal C}_{j}$ will go at most as ${\mathcal O}(1)$ while their combination ${\mathcal C}_{ij}-{\mathcal C}_{i} {\mathcal C}_{j}$ will vanish at least as ${\mathcal O}(k^2 \Delta \eta^2)$ in the IR limit $k \Delta \eta \ll 1$.  As a consequence, the infrared part of the terms  in the induced backreaction and in the dispersion  will give a subleading contribution, and we can safely fix our infrared cut-off to be $k= H_0$ -- i.e. we can limit ourselves to sub-horizon modes -- without affecting  the final result.

\begin{figure}[b!]
\centering
\includegraphics[width=15cm]{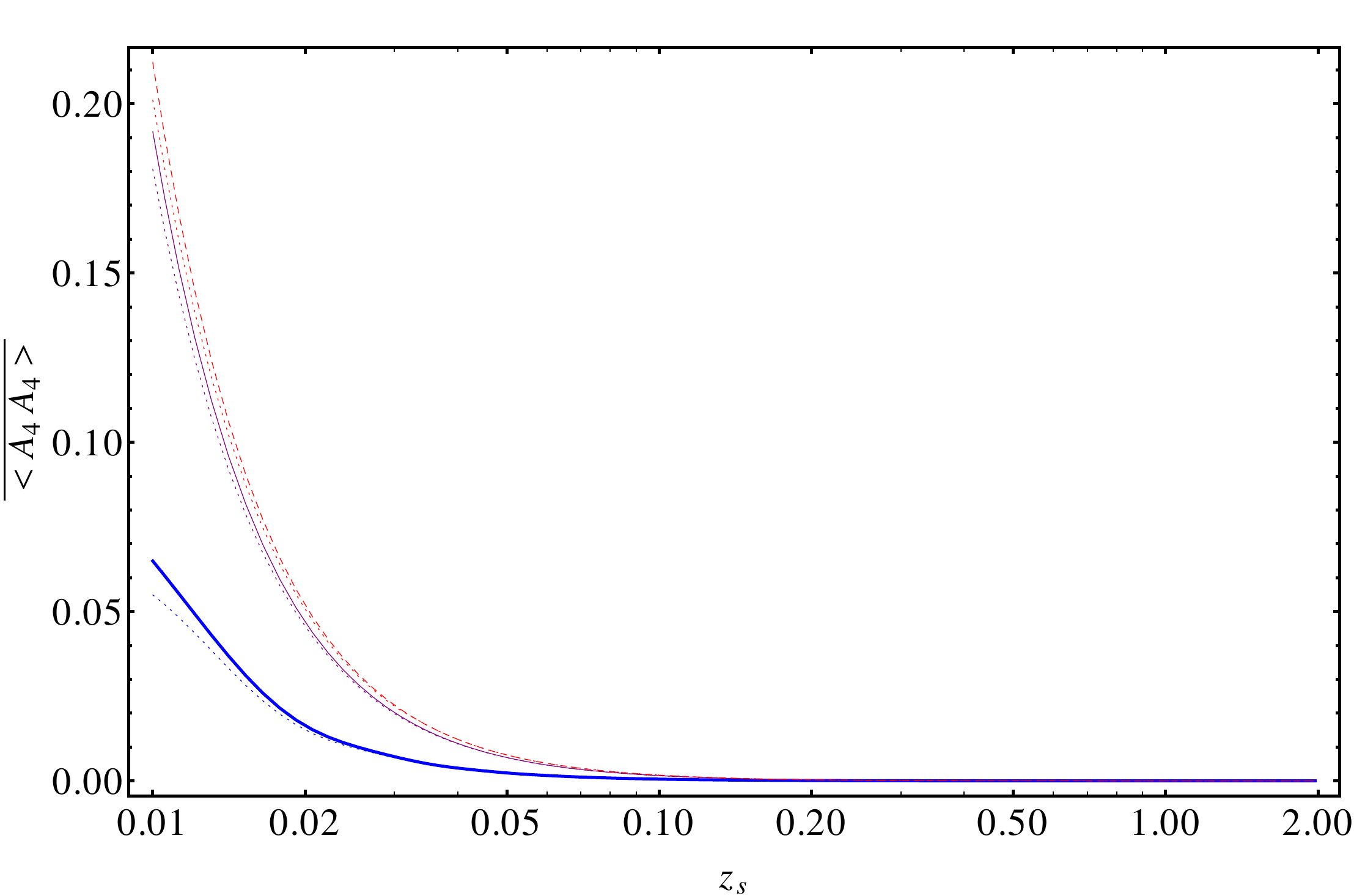}
\centering
\caption{The result of the numerical integration of Eq. (\ref{aiaj}) for  
$\overline{\lla A_4 A_4 \rra}$ is plotted as a function of $z_s$ for three different values of the UV cut-off: $k= 0.1 \,{\rm Mpc}^{-1}$ (thick blue line), $k=1  \,{\rm Mpc}^{-1}$ (thin purple line), $k=+\infty$ (dashed red line). Also shown (by the corresponding dotted curves) is the plot of the contribution to IBR$_2$  $ \overline{\langle A_4A_4 \rangle}  -\overline{\langle A_4 \rangle \langle  A_4 \rangle}$.}
\label{Fig1}
\end{figure}

\begin{figure}[ht!]
\centering
\includegraphics[width=15cm]{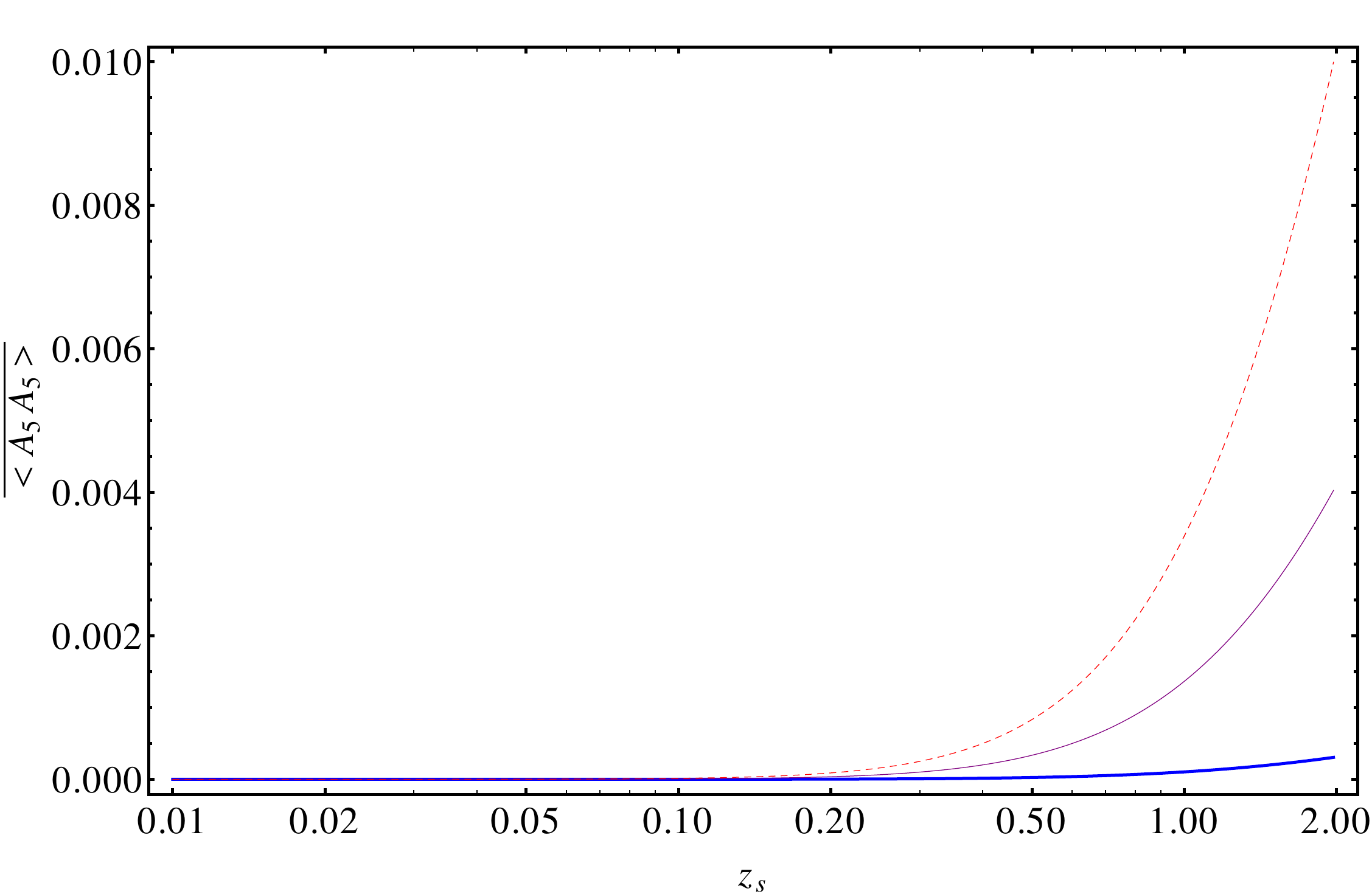}
\centering
\caption{The result of the numerical integration of Eq. (\ref{aiaj}) for 
$\overline{\lla A_5 A_5 \rra}$ is plotted as a function of $z_s$ for three different values of the UV cut-off: $k= 0.1 \,{\rm Mpc}^{-1}$ (thick blue line), $k=1  \,{\rm Mpc}^{-1}$ (thin purple line), $k=+\infty$ (dashed red line).}
\label{Fig2}
\end{figure}

Furthermore, as can be seen in Table \ref{Tab1} and \ref{Tab2}, all the coefficients ${\mathcal C}_{ij}$ and ${\mathcal C}_{i} {\mathcal C}_{j}$ will go at most as ${\mathcal O}(k^3 \Delta \eta^3)$ in the UV limit $k \Delta \eta \gg 1$. As a consequence, all the integrals involved in the induced backreaction and in the dispersion will be UV-convergent and their main contribution will come from the range $1/\Delta \eta \ll k \leq 2.5 \, {\rm Mpc}^{-1}$. In this range, some of the backreaction coefficients grow as positive power of $k$ while the transfer function is not yet decreasing as 
$\log k/k^2$. In particular, there are only two leading contributions corresponding to the integrals controlled by the coefficients ${\mathcal C}_{44}$ and ${\mathcal C}_{55}$ (this is the reason why the dispersion of the angular average of $d_L$, (\ref{varp}), is smaller than the one of $d_L$ itself (\ref{vargen2nd})). For such coefficients we have, in fact, the following behaviour for $k \Delta \eta \gg 1$:
\begin{eqnarray}
{\mathcal C}_{44} &\simeq & \left(1 - \frac{1}{{\mathcal H}_s\Delta \eta}\right)^2  
(f_0^2 + f_s^2) \frac{k^2}{3} ,
\\
{\mathcal C}_{55} &\simeq & \frac{k^3 \Delta \eta^3}{15} {\rm SinInt}(k \Delta \eta)\,.
\end{eqnarray}
From the  above expressions it is easy to understand that
$\overline{\lla A_4 A_4 \rra}$ will give the largest contribution for $z_s\ll 1$ (see Fig. \ref{Fig1}), while $\overline{\lla A_5 A_5 \rra}$ will give 
the largest contribution for $z_s\gg 1$ (see Fig. \ref{Fig2}).

It is also clear that both the induced backreaction IBR$_2$ and the dispersion depend in principle on the UV cut-off $k_{UV}$ eventually used to evaluate the integrals. On the other hand, when $k_{UV}$ is taken inside the regime where  the spectrum goes like $(\log k)^2/k^4$, the dependence on the cut-off will not be too strong. In particular the leading contribution to IBR$_2$, given by $2 \overline{\lla A_4 A_4 \rra}$, depends very weakly on the particular value of $k_{UV}$.  The leading contribution to the dispersion (\ref{BRVarTh}), controlled by $\overline{\lla A_5 A_5 \rra}$, has a somewhat stronger dependence on $k_{UV}$ (because of the extra power of $k$  in the integrand of $\overline{\lla A_5 A_5 \rra}$). The numerical integrations of $\overline{\lla A_4 A_4 \rra}$ and $\overline{\lla A_5 A_5 \rra}$ are presented in Figs. \ref{Fig1} and \ref{Fig2}, where we illustrate the magnitude of the backreaction effects as a function of the redshift (in the range of values relevant to supernovae observations), and as a function of its dependence on  the  cut-off (ranging from $k_{UV}=0.1 \,{\rm Mpc}^{-1}$ to $k_{UV}=+\infty$). We should emphasize that the two abovementioned contributions
have a clear and distinct physical meaning in the Newtonian gauge. Going back to their explicit expressions it appears that $\overline{\lla A_4 A_4 \rra}$ represents a
Doppler effect  while $\overline{\lla A_5 A_5 \rra}$ is associated with 
lensing, both in qualitative agreement with previous claims in the literature \cite{Flanagan, Bonvin}. Note that the latter term can only appear  in the
variance, because it contributes to $\sigma_1^2$, while
 $\overline{\lla A_4 A_4 \rra}$ appears also in
IBR$_2$ since $A_4$ is  present in both $\mu_1$ and $\sigma_1$.
In Fig. 1 we have also plotted the full  contribution to IBR$_2$, namely 
$ \overline{\langle A_4A_4 \rangle}  -\overline{\langle A_4 \rangle \langle  A_4 \rangle}$ (see Eq. (\ref{BR2backreaction})): we can see that the difference from the behavior of the $ \overline{\langle A_4A_4 \rangle}$ term is very small, with no qualitative effect on our discussion.

It is important to stress that, although all considered backreaction contributions are (at any redshift) UV-finite, they can induce relatively big effects at the distance scales  relevant for supernovae observations (see Figs.  \ref{Fig1}  and \ref{Fig2}), provided the given spectrum is extrapolated to sufficiently large values of $k$.
We also stress that the decoupling of the small-distance (i.e. high-$k$) scales from the considered large-scale backreaction is due to the efficient suppression of the linear perturbation modes inside the horizon, an effect well described by the transfer function (\ref{EH97}).

Let us now sum up all contributions to the induced backreaction (\ref{BR2backreaction}) and to the dispersion (\ref{BRVarTh}), and compare the results for $\overline{\langle d_L \re} \pm d_L^{CDM}\sqrt{\overline{\langle \sg_1^2 \re}}$ with the homogeneous luminosity-distance of a pure CDM model and of a successful  $\La$CDM model. We will include into $\overline{\langle d_L \re}$ only the IBR$_2$ contribution (\ref{ibr2}), namely we will set $\overline{\langle d_L \re}= d_L^{CDM}(1+ {\rm IBR}_2)$. It is clear that a full computation should include additional contributions, of the same order as IBR$_2$ or even larger, arising from second-order perturbations of $d_L$ (we are referring to the term called $\overline{\langle \sg_2 \re}$ in Eq. (\ref{EnsAvExp})). Nonetheless, we believe that (modulo cancellations) our computation may estimate a reliable ``lower limit" on the strength of the possible corrections to the luminosity-redshift relation in the context of our  inhomogeneous geometry. 

\begin{figure}[h!]
\centering
\includegraphics[width=14.5cm]{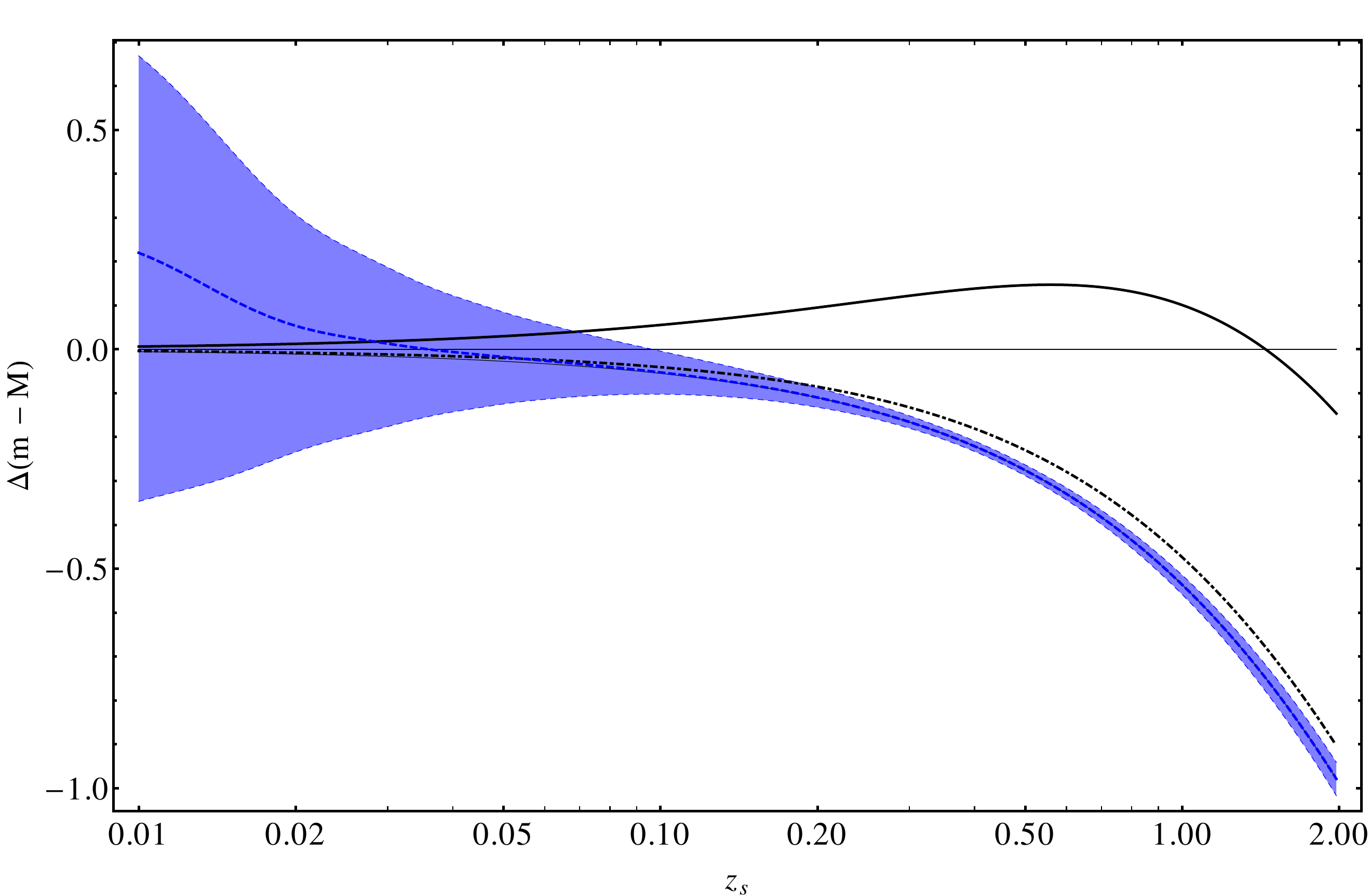}
\centering
\caption{The distance-modulus difference of Eq. (\ref{modulus}) is plotted for a pure CDM model (thin line), for a CDM model including the contribution of IBR$_2$ (dashed blue line) plus/minus the dispersion (coloured region), and for a $\Lambda$CDM model with $\Om_\Lambda=0.73$ (thick line) and $\Om_\Lambda=0.1$ (dashed-dot thick line). We have used for all backreaction integrals the cut-off $k=0.1 \,{\rm Mpc}^{-1}$.}
\label{Fig3}
\end{figure}

\begin{figure}[ht!]
\centering
\includegraphics[width=14.5cm]{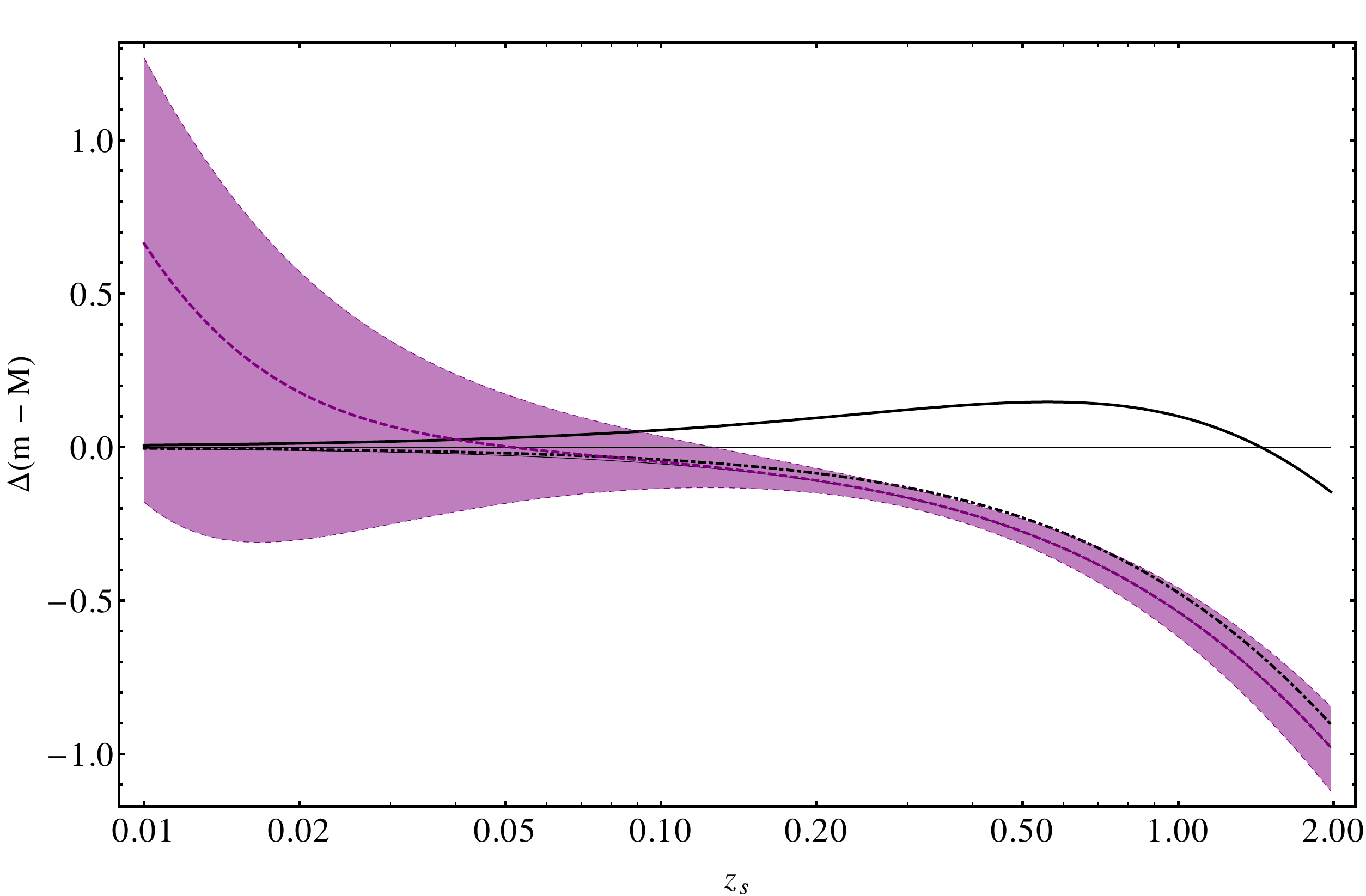}
\centering
\caption{The  distance-modulus difference  of Eq. (\ref{modulus}) is plotted for a pure CDM model (thin line), for a CDM model including the contribution of IBR$_2$ (dashed blue line) plus/minus the dispersion (coloured region), and for a $\Lambda$CDM model with $\Om_\Lambda=0.73$ (thick line) and $\Om_\Lambda=0.1$ (dashed-dot thick line). We have used for all backreaction integrals the cut-off $k=1 \,{\rm Mpc}^{-1}$.}
\label{Fig4} 
\end{figure}

The comparison between the homogeneous and inhomogeneous (averaged) values of $d_L$ can be conveniently illustrated by plotting the so-called distance modulus $m-M= 5 \log_{10}d_L$ or, even better, by plotting the difference between the distance modulus of the considered model and that of a flat, linearly expanding Milne-type geometry, used as reference value (see e.g. \cite{MG}). In such a case we can plot, in particular, the following quantity:
\be
\Delta(m-M)=5 \log_{10} \left[ \overline{\langle d_L \re} \right] - 5 \log_{10}\left[\frac{(2+z_s)z_s}{2 H_0}\right]. \label{modulus}
\ee
The results are illustrated in Fig. \ref{Fig3} for the case of a  cut-off 
$k_{UV}=0.1 \,{\rm Mpc}^{-1}$, and in Fig. \ref{Fig4} for $k_{UV}=1 \,{\rm Mpc}^{-1}$. The averaged luminosity-redshift relation of our inhomogeneous model is compared, in particular, with that of a pure CDM model and with that of a $\La$CDM model with $\Om_\La=0.1$ and with $\Om_\La= 0.73$. We have also explicitly shown the expected dispersion around the averaged result, by plotting the curves corresponding to $\overline{\langle d_L \re} \pm d_L^{CDM}\sqrt{\overline{\langle \sg_1^2 \re}}$ (bounding the coloured areas appearing in the figures). 


Let us stress again that the choice of the cut-off may affect (even if not dramatically) the final result when the values of $k_{UV}$ are varying in the range $(0.1 - 1) \,{\rm Mpc}^{-1}$, while the precise choice becomes less important at higher values of $k_{UV}$. We should recall, however, that the inhomogeneous model adopted in this paper is fully under control only in the linear perturbative regime, and that the spectrum cannot be extrapolated at scales higher than about $k\sim1 \,{\rm Mpc}^{-1}$ without taking into account the complicated effects of its non-linear dynamical evolution. 
The approximate coincidence between the above limiting value of $k_{UV}$ and the scale marking the beginning of the non-linear perturbative regime is not only a particular consequence of the transfer function adopted in this paper (see Eq. (\ref{EH97})), but also an avoidable property of realistic non-linear perturbations spectra (see e.g. \cite{smith}).

As clearly shown by the two figures, the corrections induced by IBR$_2$ on the luminosity distance of a homogeneous CDM model, 
even taking into account the expected dispersion of values around $ \overline{\langle d_L \re}$, cannot be used to successfully simulate realistic dark-energy effects. In the figures we have also plotted, for illustrative purposes,  an example of standard $\La$CDM model with $\Om_\La=0.1$, which seems to be compatible (within the allowed region defined by the dispersion) with the prediction of our simple CDM+IBR$_2$ model, at least for sufficiently high values of the cut-off scale (see in particular Fig. \ref{Fig4}).  However, we warn the reader that it would be wrong to conclude that our backreaction can mimic a fraction of dark energy of the order of $\Om_\La \sim 0.1$, because a similar conclusion might be reliably reached only by averaging inhomogeneities on a background which already includes a significant amount of dark energy from the beginning. 

Let us conclude this section with some important comments. First, as already stressed, a consistent second-order computation of the backreaction should include the contribution of $\overline{\langle \sg_2 \re}$. This requires a full treatment of metric perturbations and coordinate transformations up to second order (work is in progress on this point \cite{work}).  Nevertheless, we can easily see from Eqs.(\ref{General_Eq_dA}) and (\ref{dLaverageexact}) that  $\overline{\lla \sigma_2 \rra}$  contains, among others,  contributions of the type $\overline{\lla A_5 A_5 \rra}$ already computed in this paper. The  behaviour of this term, in the asymptotic regime $k\Delta \eta \gg 1$, is very different from the behaviour of  terms like $\overline{\lla A_4 A_4 \rra}$ which give the leading contribution to IBR$_2$: the contribution of  $\overline{\lla A_5 A_5 \rra}$, in particular, grows at large redshifts, as illustrated in Fig. \ref{Fig2}. 
Using our results for $\overline{\langle A_5 A_5 \rangle}$ -- and barring cancellations -- we can obtain, for instance, the following numerical estimates:
with an UV cutoff $k_{UV}=1 \,{\rm Mpc}^{-1}$ we expect  $|\overline{\langle \sg_2 \re}|  \gaq 1.5 \times 10^{-3}$ at $z=1$  and 
$|\overline{\langle \sg_2 \re}|  \gaq 4\times 10^{-3}$ at
 $z=2$. These  expectations do not change much if we increase further the UV cutoff: they  only go up by about a factor two  even if we send  $k_{UV}$ arbitrarily high (at least within our form of the power spectrum).  

These examples suggest that a full computation of  $\overline{\langle \sg_2 \re}$ could strongly enhance the overall backreaction effects at large $z_s$, with respect to the effects due to IBR$_2$ discussed previously. Hence a full second-order computation, possibly joined to a reliable estimate of contributions from the non-linear regime\footnote{The importance of such a non linear regime was also recently underlined in \cite{CEFMUU}, following a different approach to describe the impact of inhomogeneities on the supernovae observations.}, appears to be necessary before firm conclusions on the correct interpretation of the data can be drawn. Also, the different behaviour of the different backreaction contributions, at small $z_s$ and large $z_s$, could represent an important signature to distinguish the  effects due to averaged inhomogeneities from the more conventional dynamical effects of homogeneous dark energy sources.

The second comment  is that, although a reliable estimate of the full backreaction on the averaged luminosity distance requires a full second-order calculation, some suitable linear combinations of averages of different powers of  $ d_L$ only depend on the first-order quantity $\sigma_1$ (defined by the expansion of $d_L$). As an example, one can show that the following equality holds at second order for any value of the real parameter $\alpha$:
\be
 \overline{\lla \left(d_L/d_L^{FLRW}\right)^{\alpha} \rra}  - \alpha  \overline{\lla d_L/d_L^{FLRW} \rra} = 1-\alpha + \frac{\alpha (\alpha -1)}{2} \overline{\langle \sigma_1^2 \rangle}.
\ee
This quantity can be plotted for a given inhomogeneous model, and compared with  its (deterministic) value in a $\Lambda$CDM model, for various values of  $\alpha$. The result is  that the two models disagree for generic $\alpha$, leading to the conclusion that realistic  inhomogeneities added to CDM lead to a model that can be distinguished, in principle,  from the conventional $\Lambda$CDM scenario. In practice, however, we only have a single quantity measured by the supernovae experiments (basically the received  flux of radiation), and one cannot  exclude that the two models happen to give the same result for that particular observable.


\section{Conclusions}
\label{Sec7}
\setcounter{equation}{0}

Let us  briefly summarize the main ideas and results of this work.
Using a previously introduced  gauge invariant light-cone averaging prescription, as well as an adapted coordinate system, we were able to write down an exact expression for the averaged luminosity distance $d_L(z)$ (or any function of it) as a function of the redshift $z$.
In principle such an expression can be used to study the effect of inhomogeneities even outside the domain of cosmological perturbation theory. In this paper we have only attempted a first study of the leading-order effect generated by  stochastic perturbations of the type produced by inflation on top of a pure CDM model.

We have then been able to separate two distinct contributions to the backreaction. The first, that we called induced backreaction (IBR), is sensitive to our specific averaging prescription and originates from correlations in the fluctuations of the luminosity distance and those in the integration measure. As a consequence, IBR can be computed to second order (giving what we have called IBR$_2$) by using results from linear perturbation theory. The second contribution is insensitive to the averaging prescription but needs a full computation of second order perturbations in the luminosity distance itself, something that we are leaving to future work. Linear perturbation theory also allows to compute the expected variance on the (angular and ensemble) average of $d_L(z)$.

Our final integrals for these quantities are nicely behaved both in the infrared (showing explicitly that perturbations on scales much larger than the source's distance do not contribute) and  in the ultraviolet, provided short-scale perturbations behave roughly as in simple, realistic models for the matter power spectrum. This does not mean, however, that the final result is insensitive to the detailed structure of inhomogeneities in the non-linear (or even non-perturbative) regime.

Although the power spectrum has an overall normalization of order $10^{-9}$ (from CMB data),  it induces a backreaction which is strongly enhanced by phase-space factors of order $(k_*/H_0)^p$ (with $p$ a power that can be as big as $2$ or $3$), where $k_*$ is some typical scale appearing in the power spectrum.  These enhancement factors can bring the IBR effects many orders of magnitude higher. However,
if we use a rather low momentum cutoff (by insisting on staying inside the linear regime), IBR$_2$ effects turn out to be way too small to mimic an appreciable cosmological constant (see for instance Fig. {\ref{Fig3}). The total backreaction  could instead be larger, but a reliable determination of its magnitude depends on being able to carry out the full second-order calculation mentioned above, as well as on having some knowledge about the non-perturbative (short-distance)  contribution to $d_L$.

We have also noticed that the expected (angular) dispersion of $d_L$ is quite large (see Figs. {\ref{Fig3} and {\ref{Fig4}), particularly at small and at high $z_s$ (with a minimum in the region of intermediate $z_s$, i.e. $z_s$ = 0.3-0.6), something that one should be able to check quite precisely once more data become available.

We note that the dominant contribution to the dispersion at small $z$ comes from the $\overline{\langle A_4 A_4\rangle}$ term, and is therefore associated with ``Doppler-type" contributions, since the quantity $I_r$ is related to the source and observer velocities. This is in agreement with the claims made in \cite{Flanagan}. By contrast, the main contribution at large $z$ comes from  
$\overline{\langle A_5 A_5\rangle}$, i.e. from terms usually referred to as ``lensing" contributions. Again, this is in agreement with the literature (see  \cite{Bonvin}).

The size and the nice IR and UV behaviour of our gauge invariant quantities can be contrasted with what happens if one approaches the backreaction problem using averaging on  spacelike hypersurfaces, as in \cite{CU,Kolb}. In that case the calculation of the average expansion rate is UV convergent and practically 
cut-off independent. In a reasonable range of cut-off values it gives a 
backreaction of order $10^{-5}$ for the concordance
model. However, for other quantities such as the variance of
the expansion rate or the deceleration parameter,  terms like
$\sim (\partial^2 \Psi)^2$ appear, giving UV-divergent integrals and a backreaction of $O(1)$ for relatively low cut-off. In our case, considering a UV cut-off equal to $1\, {\rm Mpc}^{-1}$ we obtain a backreaction effect (estimated from $\overline{\lla A_5 A_5 \rra}$) of the order of $10^{-3}$ for $z\sim1$, namely 2 orders of magnitude bigger than the result presented in  \cite{CU, Kolb}.
Furthermore, there is no qualitative difference between the calculation of the induced backreaction and the one of the variance. They are all given by nice IR and UV-convergent integrals which never produce numbers of $O(1)$ even for very large momentum cutoffs.

Finally, by suitably combining averages of different  functions of $d_L$, we can again obtain results that, like the variance, only depend  on linear perturbation theory. They show that, in principle, an inhomogeneous CDM model can be neatly distinguished from a homogeneous $\Lambda$CDM model. It remains to be seen whether any such combination is accessible to observations. Nonetheless, in the light of this last observation, it would appear quite unlikely that one will  be able to fully account  for the supernovae data in terms of inhomogeneity effects. Rather, depending on the full contribution 
of the second-order terms and, possibly, on the one from the highly non-linear regime, the effect of inhomogeneities could be relevant for future precise determinations of the critical fraction and equation of state of dark energy. In this spirit we plan to extend the calculations presented here to the (only slightly more complicated) case of generic $\Lambda$CDM models.


\section*{ACKNOWLEDGMENTS}
We wish to thank Luca Amendola, Ramy Brustein, Thomas Buchert, Chris Clarkson, Ruth Durrer, 
Valerio Marra, Misao Sasaki, Dominik
Schwarz and  Luigi Tedesco  for useful comments and discussions.
IBD and MG are grateful to the Coll\`ege de France for hospitality and financial support, and IBD also gratefully acknowledges partial financial support from  G.R.A.M. funds.
The research of IBD at Perimeter Institute is supported by the
Government of Canada through Industry Canada and by the Province of Ontario through the Ministry of Research \& Innovation. IBD is also supported in part by funding from the
Canadian Institute for Advanced Research. GV would like to thank the Yukawa Institute for Theoretical Physics (YITP) in Kyoto, Japan, for hospitality during part of this work and the JSPS Invitation Fellowship Program for
Research in Japan NO. S-11136 for supporting that visit.


\begin{appendix}
\renewcommand{\theequation}{A.\arabic{equation}}
\setcounter{equation}{0}
\section*{Appendix A. Other examples of backreaction integrals}

Here we present a detailed computation of some of the terms contributing to the induced backreaction and to the dispersion associated to the light-cone average of $d_L(z_s, \theta^a)$. 
Such computations also define the corresponding spectral coefficients appearing in Table I and Table II. We will start with  $\overline {\lla A_2 A_5 \rra}$ and $\overline {\lla A_2 \rra \lla A_5 \rra}$ as  illustrative examples. We will then consider the two leading contributions 
given by $\overline{\lla A_4 A_4 \rra}$ and $\overline{\lla A_5 A_5 \rra}$ (and, for completeness, the associated terms $\overline{\lla A_4 \rra \lla A_4 \rra}$ and $\overline{\lla A_5 \rra \lla A_5 \rra}$).

Using equations (\ref{J2_Equation}), (\ref{Psi_average_Equation}),  \rref{defA}, and working in the hypothesis of time-independent  $\Psi_k$, we obtain:
\bea
\overline {\lla A_2 A_5 \rra} &=& \int \frac{d^3 k ~d^3 k'}{(2 \pi)^{3}} \overline{E(\vec{k}) E(\vec{k'})} \int \frac{d^2 \Omega}{4 \pi} \left[ \frac{2}{\Delta\eta} \int_{\eta_s}^{\eta_0} d\eta' ~ \Psi_k \, e^{i (\eta_0-\eta') \vec{k}\cdot \hat{x}} \right] \left[ - \int_{\eta_s}^{\eta_0} \frac{d\eta''}{\Delta \eta} \frac{\eta'' - \eta_s}{\eta_0 - \eta''} ~ \Delta_2 \left(\Psi_{k'} \,
e^{i (\eta_0-\eta'')  \vec{k'}\cdot\hat{x}}\right) \right] \nonumber \\
&=& \int \frac{d^3 k}{(2 \pi)^{3}} ~ |\Psi_k|^2 \int_{-1}^{1} \frac{d(\cos\theta)}{2} \left[ \frac{2}{\Delta\eta} \int_{\eta_s}^{\eta_0} d\eta' ~ e^{i k (\eta_0 - \eta') \cos\theta} \right] \nonumber \\
& & \times \left[ \frac{1}{\Delta \eta} \int_{\eta_s}^{\eta_0} d\eta'' \frac{\eta'' - \eta_s}{\eta_0 - \eta''} \left( k^2 (\eta_0 - \eta'')^2 \sin^2 \theta - 2 i k (\eta_0 - \eta'') \cos\theta \right) e^{- i k (\eta_0 - \eta'') \cos\theta} \right] \nonumber \\
&=&  \int_0^{\infty} \frac{d k}{k} ~ P_{\Psi}(k) \cdot \frac{2}{3 (k \Delta \eta)^2} 
\left[-4 + (4 + (k \Delta \eta)^2)\cos(k \Delta \eta) + k \Delta \eta \sin(k \Delta \eta) + (k \Delta \eta)^3 {\rm SinInt}(k \Delta \eta) \right];
\label{A1}
\eea

\bea
\overline {\lla A_2 \rra \lla A_5 \rra} &=& \int \frac{d^3 k ~d^3 k'}{(2 \pi)^{3}} \overline{E(\vec{k}) E(\vec{k'})} \left[ \int \frac{d^2 \Omega}{4 \pi} \frac{2}{\Delta\eta} \int_{\eta_s}^{\eta_0} d\eta' ~ \Psi_k \, e^{i (\eta0 - \eta') \vec{k}\cdot \hat{x}} \right] 
\nonumber \\
& & \times
\left[ - \int \frac{d^2 \Omega'}{4 \pi} \int_{\eta_s}^{\eta_0} \frac{d\eta''}{\Delta \eta} \frac{\eta'' - \eta_s}{\eta_0 - \eta''} ~ \Delta_2 \left(\Psi_{k'} \, e^{i (\eta_0 - \eta'')  \vec{k'}\cdot\hat{x}}\right) \right] \nonumber \\
&=&  \int \frac{d^3 k}{(2 \pi)^{3}} ~ |\Psi_k|^2 \left[ \int_{-1}^{1} \frac{d(\cos\theta)}{2} \frac{2}{\Delta\eta} \int_{\eta_s}^{\eta_0} d\eta' ~ e^{i k (\eta_0 - \eta') \cos\theta} \right] \nonumber \\
& & \times\left[ \int_{-1}^{1} \frac{d(\cos\theta')}{2} \frac{1}{\Delta \eta} \int_{\eta_s}^{\eta_0} d\eta'' \frac{\eta'' - \eta_s}{\eta_0 - \eta''} \left( k^2 (\eta_0 - \eta'')^2 \sin^2 \theta' - 2 i k (\eta_0 - \eta'') \cos\theta' \right) e^{- i k (\eta_0 - \eta'') \cos\theta'} \right] \nonumber \\
&=& \int_0^{\infty} \frac{d k}{k} ~ P_{\Psi}(k) \cdot \left[ \frac{2}{k\Delta \eta} {\rm SinInt}(k\Delta \eta) \right] \times 0 ~ = ~ 0. 
\label{A2}
\eea
In the second lines of Eqs. (\ref{A1}) and (\ref{A2}) we have applied Eq.  
\rref{PropRandomVar} to remove the integration over $\vec{k'}$, and we have used the isotropy of the scalar product $\vec{k}\cdot \hat{x}$. 

We move now to the terms representing the two leading contributions to IBR$_2$ and to the dispersion.
Starting from Eq. \rref{Ir}, and using the same hypotheses as in the previous calculations, we have
\bea
I_r &=& \int_{\eta_{in}}^{\eta_s} d\eta' \frac{a(\eta')}{a(\eta_s)} \pa_r \Psi(\eta',\eta_0-\eta_s,\theta^a)- \int_{\eta_{in}}^{\eta_0} d\eta' \frac{a(\eta')}{a(\eta_0)} \pa_r   \Psi(\eta', 0, \theta^a) \nonumber \\
&=& \int \frac{d^3 k}{(2 \pi)^{3/2}} \left\{ \int_{\eta_{in}}^{\eta_s} d\eta' \frac{\eta'^2}{\eta_s^2} (i k \cos\theta) \Psi_k \, e^{i \Delta \eta \vec{k}\cdot \hat{x}} - \int_{\eta_{in}}^{\eta_0} d\eta' \frac{\eta'^2}{\eta_0^2} (i k \cos\theta) \Psi_k \right\} \nonumber \\
&=& \int \frac{d^3 k}{(2 \pi)^{3/2}} \Psi_k \cdot \left( f_s e^{i k \Delta\eta \cos\theta} - f_0 \right) (i k \cos\theta) \,.
\eea
Then, according to the definitions (\ref{defA}): 
\be 
\overline {\lla A_4 A_4 \rra} = \left(1  -  \frac{1}{{\mathcal H}_s\Delta \eta}\right)^2 \overline {\lla I_r I_r \rra},
\ee
where
\bea
\overline {\lla I_r I_r \rra} &=& \int \frac{d^3 k}{(2 \pi)^{3}} ~ |\Psi_k|^2 \int_{-1}^{1} \frac{d(\cos\theta)}{2} \left[ \left( f_s e^{i k \Delta\eta \cos\theta} - f_0 \right) (i k \cos\theta) \right] \left[ \left( f_s e^{-i k \Delta\eta \cos\theta} - f_0 \right) (- i k \cos\theta) \right] \nonumber \\
&=& \int_0^{\infty} \frac{d k}{k} ~ P_{\Psi}(k) \cdot \frac{1}{3 \Delta\eta^2} \left[ (f_0^2 + f_s^2) ( k \Delta\eta)^2 - 12 f_0 f_s \cos(k \Delta\eta) - 6 f_0 f_s ( -2 + (k \Delta\eta)^2)\frac{\sin(k \Delta\eta)}{k \Delta\eta} \right],
\eea
and where we have used the following exact integral result:
\be
 \int_{-1}^{1} \frac{d(\cos\theta)}{2} (\cos\theta)^2 e^{\pm i k \Delta\eta \cos\theta} = \frac{2 k \Delta\eta \cos(k \Delta\eta) + (-2 + (k \Delta\eta)^2) \sin(k \Delta\eta)}{(k \Delta\eta)^3} \,.
\ee

In the same way we obtain:
\beq
\overline {\lla A_4 \rra \lla A_4 \rra} = \left(1  -  \frac{1}{{\mathcal H}_s\Delta \eta}\right)^2 \overline {\lla I_r \rra \lla I_r \rra},
\eeq
where
\bea
\overline {\lla I_r \rra \lla I_r \rra} &=& \int \frac{d^3 k}{(2 \pi)^{3}} ~ |\Psi_k|^2 \left[ \int_{-1}^{1} \frac{d(\cos\theta)}{2} \left( f_s e^{i k \Delta\eta \cos\theta} - f_0 \right) (i k \cos\theta) \right] 
\\ && \times
\left[ \int_{-1}^{1} \frac{d(\cos\theta')}{2} \left( f_s e^{- i k \Delta\eta \cos\theta'} - f_0 \right) (- i k \cos\theta') \right] \nonumber \\
&=& \int_0^{\infty} \frac{d k}{k} ~ P_{\Psi}(k) \cdot \left[ \frac{f_s}{\Delta \eta} \left( \cos(k \Delta\eta) - \frac{\sin(k \Delta\eta)}{k \Delta\eta} \right) \right]^2,
\eea
and where we have used  the integral
\be
 \int_{-1}^{1} \frac{d(\cos\theta)}{2} ~ \cos\theta e^{\pm i k \Delta\eta \cos\theta} = \pm i \left(\frac{- k \Delta\eta \cos(k \Delta\eta) + \sin(k \Delta\eta)}{(k \Delta\eta)^2}\right) \,.
\ee

Following a similar procedure Eq.(\ref{J2_Equation}) leads us to
\bea
\overline {\lla A_5 A_5 \rra} &=& \int \frac{d^3 k}{(2 \pi)^{3}} ~ |\Psi_k|^2 \int_{-1}^{1} \frac{d(\cos\theta)}{2} \left[ \int_{\eta_s}^{\eta_0} \frac{d\eta'}{\Delta \eta} \frac{\eta' - \eta_s}{\eta_0 - \eta'} \left( k^2 (\eta_0 - \eta')^2 \sin^2 \theta + 2 i k (\eta_0 - \eta') \cos\theta \right) e^{i k (\eta_0 - \eta') \cos\theta} \right] \nonumber \\
& & \times \left[ \int_{\eta_s}^{\eta_0} \frac{d\eta''}{\Delta \eta} \frac{\eta'' - \eta_s}{\eta_0 - \eta''} \left( k^2 (\eta_0 - \eta'')^2 \sin^2 \theta - 2 i k (\eta_0 - \eta'') \cos\theta \right) e^{- i k (\eta_0 - \eta'') \cos\theta} \right] \nonumber \\
&=& \int_0^{\infty} \frac{d k}{k} ~ P_{\Psi}(k) \cdot \frac{1}{15 (k \Delta\eta)^2} \left[-24 + 20 (k \Delta\eta)^2 + (24 - 2 (k \Delta\eta)^2 + (k \Delta\eta)^4) \cos(k \Delta\eta) \right. \nonumber \\
& & \left. - 6 k \Delta\eta \sin(k \Delta\eta) + (k \Delta\eta)^3 \sin(k \Delta\eta) + (k \Delta\eta)^5 {\rm SinInt}(k \Delta\eta) \right]\,.
\eea
Note that, for the same reason for which  $\overline {\lla A_2 \rra \lla A_5 \rra} = 0$, i.e. for the fact that the integrand of $A_5$ gives zero when averaged over the angles, we finally also obtain $\overline {\lla A_5 \rra \lla A_5 \rra} = 0$.

\end{appendix}


\end{document}